\documentclass[nofootinbib,reprint]{revtex4-1}
\usepackage{graphicx,color}
\usepackage{amsmath, amssymb, bm}

\usepackage{tikz}

\begin{document}

\title{ 
Bi-local fields interacting with a constant electric field and related problems including the Schwinger effect}

\author{Kenichi Aouda${}^{1,*}$, Naohiro Kanda${}^{1,\dag}$, Shigefumi Naka${}^{1,\ddag}$}
 \author{Haruki Toyoda${}^{2,\S}$}
\affiliation{${}^1$Department of Physics, College of Science and Technology, Nihon University, Tokyo 101-8308, Japan \\ ${}^2$Junior College, Funabashi Campus, Nihon University, Chiba  274-8501, Japan}


\email{\!aoda.kenichi20@nihon-u.ac.jp \\ \hspace{-2mm}\!${}^\dag$\,kanda.naohiro20@nihon-u.ac.jp \\ \!\hspace{-2mm}${}^\ddag\,$naka@phys.cst.nihon-u.ac.jp \\ \!\hspace{-2mm}${}^\S$\,toyoda.haruki@nihon-u.ac.jp}




\begin{abstract}
\begin{center}
\end{center}
The bi-local fields are the quantum fields of two-particle systems, the bi-local, systems, bounded by relativistic potentials. Since those form constrained dynamical systems, it is limited to introduce the interactions of the bi-local fields with other fields. In this paper, the interaction between the bi-local fields and a constant electric field $E$ is studied with consideration for the consistency of constraints. Then, we evaluate the Schwinger effect for the bi-local systems, which gives the  pair production probability of the bound states as a function of the charges of respective particles and the coupling constant in the binding potential. Through this analysis,  we also discuss the possibility for the dissociation of bi-local systems due to the electric field.

\end{abstract}



\maketitle

\section{Introduction}

The bi-local fields are known to be an original form of non-local fields extended in spacetime proposed by H. Yukawa\cite{Yukawa-1}-\cite{Yukawa-3}. In the beginning, this theory had been expected as a drastic approach to save the problems in local field theories such as the problems of divergence and those of unified description for many elementary particles, etc. According to the development of this theory, however,  the bi-local fields established the standpoint as an effectiver approach to the relativistic two-body bound systems\cite{Takabayasi} such as mesons in QCD. The resultant formulation of bi-local systems, the classical counterparts of the bi-local fields, had a similar structure to the string models\cite{BL-Review}. 

The bi-local systems have a structure of constrained dynamical systems based on the reparameterization invariance of the time parameters of constituent particles. The structure of constrained dynamics is simpler than that of the string model; however, because of this reason, it was not easy to formulate the interaction of the bi-local systems with other fields. In the attempts of those interacting bi-local systems, the scattering amplitudes\cite{Tanaka}, the form factor by external fields\cite{Namiki} and so on were studied. Through the study on a three vertex function of the bi-local fields, a prototype of the Witten type of vertex function in the string model was also proposed\cite{Goto-Naka}. We also point out the recent viewpoint relating the bi-local fields and the collective bi-local fields out of a higher-spin gauge theory in AdS spacetime\cite{AdS-BL}.

In this paper, we study the bi-local systems of two particles carrying the charges $g_{(1)}(>0)$ and $g_{(2)}(<0)$ under a constant electric field. We there take notice of that this type of interaction is relevant to the problem of dissociation and pair production of the bi-local systems by the electric field. In the next section, we review the formulation of free bi-local systems bounded by a relativistic Hooke type of potential.

In Sec.3, a formulation of a bi-local system under a constant electric field is given. The gauge structure of the bi-local system is represented by two constraints: the on-mass shell condition corresponding to a master wave equation of this system and a supplementary condition eliminating redundant degrees of freedom. The consistency of those constraints are spoiled by the presence of the electric field in general; and, we try the modification of those constraints so as to recover the consistency. 

In  Sec.4, the structure of the physical states satisfying the mass shell condition with the supplementary condition are studied. Therein, the dynamical variables of the system are resolved into the components lie in $\|$ and $\perp$ planes, where the $\|$ is the plane spanned by the direction of time ($\bm{e}^0$) and that of electric field $E\bm{e}^1$, and $\perp$ is the $(\bm{e}^2,\bm{e}^3$) plane orthogonal to $\|$
\footnote{We write the basis of 4-dimensional spacetime as $\bm{e}^\mu\,(\mu=0,1,2,3)$, to which diag$(\eta^{\mu\nu}=\bm{e}^\mu\cdot\bm{e}^\nu)=(-+++)$.}
. In terms of those components, we show the way to construct a set of consistent constraints under the presence of $E\neq 0$. Under those constraints, in Sec.5, the structure of the physical states are studied in view of that the $\|$ components of modified center of mass momenta in the mass-shell condition appear in a form of Hamiltonian for a repulsive harmonic oscillator\cite{RHO-PRD}.

Within the physical states, the ground state $|0_g^{\|}\rangle$ for the internal variables in the $\|$ plane is particularly interesting, since it depends on the charges of respective particles $g_{(i)}(i=1,2)$ and on $E$. Furthermore, if the bi-local systems are embedded in the two-dimensional $(\bm{e}^0,\bm{e}^1$) spacetime, the $\|$ plane, the excited  states of those systems are constructed only on the ground state $|0_g^{\|}\rangle$ .

In Sec.6, we study the transition amplitude between the ground state $|0_g^{\|}\rangle$ for $E_1=0$ and one for $E_2\neq 0$. The analysis on this transition gives us a knowledge on the stability of the system under the $E$. 

In Sec.7, the discussion on the Schwinger effect for the bi-local fields is given. The bi-local systems are neutral or charged systems according as $|g_{(1)}|=|g_{(2)}|$ or $|g_{(1)}|\neq |g_{(2)}|$ respectively. The analysis of the Schwinger effect in the case of $|g_{(1)}|=|g_{(2)}|$ will give us the knowledge on the dissociation of bound states by  the $E$.

Section\,8 is devoted to the summary of our results. In the Appendices, some mathematical problems used in the text are discussed: in particular in Appendix C,  the transition amplitudes between the ground states of different $E$'s are evaluated in detail for the analysis in Sec.6.

\section{Formulation of free bi-local fields}

A commonly used approach to the bi-local system, the two particle system bounded by a confining potential, is to start from the action
\begin{align}
S=\int d\tau \frac{1}{2}\sum_{i=1}^2 \left\{ e_{(i)}^{-1}\dot{x}_{(i)}^\mu \dot{x}_{(i)\mu}-V_{(i)}\!\left(x\right)e_{(i)} \right\}, \label{BL action}
\end{align}
where $x_{(i)}(\tau)^\mu~(i=1,2)$ are the coordinates of respective particles; and, the interaction terms $V_{(i)}(x)~(i=1,2)$ are set as functions of $x\equiv x_{(1)}-x_{(2)}$ to ensure the translation invariance of this system. The $e_{(i)}(\tau),(i=1,2)$ are einbeins, which guaranty the invariance of $S$ under the $\tau$ reparametrization.  When this invariance is consistent with the dynamics of bi-local system, the interaction via $V_{(i)}$'s can be described as an action at a distance between $x_{(i)}(\tau)\,(i=1,2)$ with the same $\tau$. 
\begin{figure}
\begin{minipage}{4cm}
\includegraphics[width=4cm]{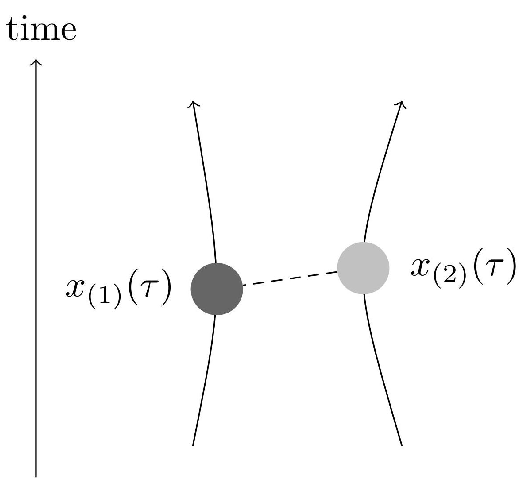}
\end{minipage}
\caption{The figure shows the world lines of a two particle system, the bi-local system, interacting each other through the confining potential $V_{(i)}(x(\tau))$ at the same time parameter $\tau$. If we regard those particles as the quark and the anti-quark, though the particle's spins are out of consideration, the bi-local system becomes a model of meson.}
\label{bi-local system}
\end{figure}
Then varying the action with respect to $e_{(i)}$, we obtain the constraints
\begin{align}
 H_{(i)}\equiv -2\frac{\delta S}{\delta e_{(i)}}=p_{(i)}^2+V_{(i)}\approx0,~(i=1,2), \label{constraint-1}
\end{align}
where $p_{(i)\mu}=\frac{\delta S}{\delta x_{(i)}^\mu}=\frac{1}{e_{(i)}}\dot{x}_{(i)\mu},(i=1,2)$ are momenta conjugate to $x_{(i)}^\mu$'s; and we have use the symbol $\approx$ for the constrained equations. The Eq.(\ref{constraint-1}) means that $m_{(i)}\equiv c^{-1}\sqrt{V_{(i)}}~(i=1,2)$ are effective masses of respective particles under the interaction. 

In what follows, we confine the discussion to the bi-local system of equal mass particles by taking the application in a latter section into account; and, we put
\begin{align}
 V_{(i)}(x)=(mc)^2+V(x)~~\left(~V(x)=\kappa^2x^2~\right) ,
\end{align}
where $c$ is the speed of light. The action (\ref{BL action}) tends to the sum of actions of free mass $m$ particles in the limit $\kappa\rightarrow 0$. In this equal mass case, one can introduce the center of mass momentum $P=p_{(1)}+p_{(2)}$ and the relative momentum $p=\frac{1}{2}(p_{(1)}-p_{(2)})$ conjugate to $X=\frac{1}{2}\left(x_{(1)}+x_{(2)}\right)$ and $x$, respectively. In terms of those variables, the constraints (\ref{constraint-1}) can be written as
\begin{align}
 H &\equiv 2(H_1+H_2)=P^2+4\left(p^2+V(x)\right)\approx 0 , \label{constraint-H} \\
 T &\equiv\frac{1}{2}(H_1-H_2)=P\cdot p\approx 0 . \label{constraint-T}
\end{align}
In the classical mechanics, because of $\{H,T\}=\kappa P\cdot x$, the equations (\ref{constraint-H}) and (\ref{constraint-T}) don't form a set of first class constraints. Even if we add $P\cdot x\approx 0$ as a secondary constraint, the algebras $\{T,P\cdot x\}=-2\kappa P^2$ leads to inconsistent secondary constraint $P^2\approx 0$.

In the quantized theories, however, this situation is successfully changed by introducing the oscillator variables $(\hat{a}^\mu,\hat{a}^{\mu\dag})$ defined by
\begin{align}
 x^\mu=\sqrt{\frac{\hbar}{2\kappa}}\left(\hat{a}^{\mu\dag}+\hat{a}^\mu\right),~\hat{p}^\mu=i\sqrt{\frac{\hbar\kappa}{2}}\left(\hat{a}^{\mu\dag}-\hat{a}^\mu \right), \label{oscillator}
\end{align}
which satisfy the commutation relation $[\hat{a}_\mu,\hat{a}_\nu^\dag]=\eta_{\mu\nu}$. In terms of those oscillator variables, Eq.(\ref{constraint-H}) can be read as the wave equation
\begin{align}
 \hat{H}|\Phi\rangle=\left(\hat{P}^2+\frac{1}{\alpha^\prime}\hat{a}^\dag\cdot\hat{a}+(m_0c)^2\right)|\Phi\rangle=0, \label{condition-H}
\end{align}
where $\alpha^\prime=\frac{1}{8\hbar\kappa}$ and $(m_0c)^2=4(mc)^2+16\hbar\kappa$. Then, Eq.(\ref{constraint-T}) should be regarded as a subsidiary condition on $\Phi$ in some way. To avoid the problem of secondary constraints, we treat Eq.(\ref{constraint-T}) as  the expectation value $\langle\Phi|\hat{T}|\Phi\rangle=0$ in analogy with the Gupta-Bleuler formalism in Q.E.D., which suggests to put
\begin{align}
 \hat{T}^{(+)}|\Phi\rangle \equiv \hat{P}\cdot \hat{a}|\Phi\rangle=0. \label{condition-T(+)}
\end{align}
Then, because of $[\hat{H},\hat{T}^{(+)}]=-\frac{1}{\alpha^\prime}\hat{T}^{(+)}$, the subsidiary condition (\ref{condition-T(+)}) is compatible with the on-mass shell condition (\ref{condition-H}). 

The states generated by $\{\hat{a}^{\mu\dag}\}$ with the ground state $\hat{a}^\mu|0\rangle=0\,(\langle 0|0\rangle=1)$ is an indefinite metric functional space  due to $[\hat{a}^0,\hat{a}^{0\dag}]=-1$. The subsidiary condition (\ref{condition-T(+)}) can remove the excitation of $\hat{a}^{0\dag}$, since Eq.(\ref{condition-T(+)}) implies $\hat{a}^{0}|\Phi\rangle=0$ in the rest frame of the center of mass momentum $P=(P^0,\bm{0})$. This also means that the particles characterized by Eqs. (\ref{condition-H}) and (\ref{condition-T(+)}) have $(\mbox{mass})^2c^2=\frac{1}{\alpha^\prime}\sum_{i=1}^3 n_i^2+(m_0c)^2~(\, n_i=0,1,2,\cdots \,)$
\footnote{
We note that the functional space characterized by Eqs.(\ref{condition-H}) and (\ref{condition-T(+)})  forms a non-unitary representation of the Lorentz group, which guarantees the manifest covariance of the formalism. In some cases, however, one may treat $\langle\Phi|T|\Phi\rangle=0$ as $T^{(-)}|\Phi\rangle \equiv P\cdot a^\dag|\Phi\rangle=0$. In this case, the operator $a^{0\dag}$ plays a role of annihilation operator; and, the positive definite metric functional space generated by the action of $(b^{\mu\dag})=(a^{0},a^{i\dag})$\, $(i=1,2,3)$ on the ground state $b^{\mu\dag}|0_b\rangle=0\,(\, \langle 0_b|0_b\rangle=1\,)$ gives rise to a unitary representation of the Lorentz group in a positive definite metric functional space. The difference of those two representations reflects in the phenomenological properties of the bi-local fields. In this paper, we require Eq.(\ref{condition-T(+)}) as the subsidiary condition on the bi-local fields to maintain the the manifest covariance of the  formalism.} It should be noticed that the condition (\ref{condition-T(+)}) does not exclude non-zero excitation $\{n_1\}$ even when the bi-local system exists in two-dimensional $(\bm{e}^0,\bm{e}^1)$ spaceteime; this fact is not trivial for the bi-local system interacting with a constant electric field $E\bm{e}^1$ as will be discussed in Sec.4.  

\section{Bi-local fields under a constant electric field}

It is not difficult to extend formally the action (\ref{BL action}) to a bi-local system interacting with external electric fields. Let $g_{(i)}\,(i=1,2)$ be the electric charges of the particle with the position variables $x_{(i)}\,(i=1,2)$  in (Fig.\ref{bi-local system}); we set $g_{(1)}\!>0$ and $g_{(2)}\!<0$ for the later analysis. Then, with the gauge potential $A^\mu$, the extended action should be
\begin{align}
\begin{split}
S=\int d\tau &\left[\sum_{i=1}^2 \frac{1}{2}\left\{ e_{(i)}^{-1}\dot{x}_{(i)}^\mu \dot{x}_{(i)\mu}-V_{(i)}(x)e_{(i)} \right\} \right. \\
 &\left.+\frac{g_{(i)}}{c}\dot{x}_{(i)}\cdot A(x_{(i)})\right], \label{E-BL action}
 \end{split}
\end{align}
from which the momenta conjugate to $x_{(i)}\,(i=1,2)$ are given as $p_{(i)}=\frac{1}{e}\dot{x}_{(i)}+\frac{g_{(i)}}{c}A(x_{(i)})\,(i=1,2)$.  Further, the constraint (\ref{constraint-1}) in this action becomes 
\begin{align}
 H_{(i)}\equiv -2\frac{\delta S}{\delta e_{(i)}}=\pi_{(i)}^2+(mc)^2+V(x)\approx 0 \label{constraint-2}
\end{align}
$(i=1,2)$, where $\pi_{(i)}=p_{(i)}-\frac{g_{(i)}}{c}A(x_{(i)})\,(i=1,2)$. From now on, we discuss the case in which a constant electric field $E$ is applied to the $\bm{e}^{1}=(0,1,0,0)$ direction; and we put the gauge potentials so that
\begin{align}
\begin{split}
 &A^0(x_{(i)}) =-\frac{1}{2}Ex_{(i)}^1, ~A^1(x_{(i)}) =-\frac{1}{2}Ex_{(i)}^0, \\
&A^2(x_{(i)}) =A^3(x_{(i)})=0.
 \end{split}
\end{align}
Then one can write
\footnote{Throughout this paper, we use the notation of decomposition for any four vector $(f^\mu)$ such as $f^\|=f_{\|}=(f^0,f^1),\, \tilde{f}^\|=\tilde{f}_\|=(f^1,f^0)$, and $f_\perp=f^\perp=(f^2,f^3)$.} 
\begin{align}
 \pi_{(i)\|}=\left(p_{(i)}+\frac{g_{(i)}E}{2c}\tilde{x}_{(i)}\right)_\|,~~\pi_{(i)\perp}=p_{(i)\perp}. \label{pi(i)}
\end{align}
Those $\pi_{(i)}'s$ can be decomposed into the center of mass and the relative components by some ways such as
\begin{align}
 \pi_{(1)}=\left(\frac{1}{2}\Pi+\pi\right)=\frac{1}{\sqrt{|\tilde{g}_{(2)}|}}\left(\frac{1}{2}\Pi_g+\pi_g\right), \\
 \pi_{(2)}=\left(\frac{1}{2}\Pi-\pi\right)=\frac{1}{\sqrt{|\tilde{g}_{(1)}|}}\left(\frac{1}{2}\Pi_g-\pi_g\right),
\end{align}
where $\tilde{g}_{(i)}=g_{(i)}/g\,(i=1,2)$ and $g=\frac{1}{2}\left(g_{(1)}-g_{(2)}\right)$. In the case of $g_{(1)}=-g_{(2)}=g(>0)$, those definitions give rise to $(\tilde{g}_{(1)},\tilde{g}_{(2)})=(1,-1)$ and $(\Pi_g,\pi_g)=(\Pi,\pi)$. We note that the $(\Pi_g,\pi_g)$ have the meaning of the center of mass and the relative components of $\pi_{(i)}$'s related to the charge distributions in the bi-local system. As needed, one can also use the relations
\begin{align}
 \Pi &=\pi_{(1)}+\pi_{(2)}=\frac{1}{\tilde{g}_*}\left(w_g\Pi_g+\Delta_g\pi_g\right), \\
 \pi &=\frac{1}{2}\left(\pi_{(1)}-\pi_{(2)}\right)=\frac{1}{\tilde{g}_*}\left(\frac{\Delta_g}{4}\Pi_g+w_g\pi_g\right) ,
 \end{align}
 where $w_g=\frac{1}{2}\left(\sqrt{|\tilde{g}_{(1)}|}+\sqrt{|\tilde{g}_{(2)}|}\right)$, $\Delta_g=\sqrt{|\tilde{g}_{(1)}|}-\sqrt{|\tilde{g}_{(2)}|}$, and $\tilde{g}_*=\sqrt{|\tilde{g}_{(1)}\tilde{g}_{(2)}|})$, which tend $(w_g,\Delta_g,\tilde{g}_*)\rightarrow (1,0,1)$ according as $|g_{(i)}|\rightarrow g$.
 
Now, for $E\ne 0$, the primary constraints (\ref{constraint-H}) and (\ref{constraint-T}) are modified so that 
  
\begin{align}
  H_E &=2\left(H_{(1)}+H_{(2)}\right)=H_{E\|}+H_{E\perp} \approx 0, \label{H_E}\\
  T_E &=\frac{1}{2}\left(H_{(1)}-H_{(2)}\right)=T_{E\|}+T_{E\perp} \approx 0. \label{T_E}
\end{align}
Here, the ($H_{E\perp},T_{E\perp}$) are the parts of the primary constraints  consisting of  $\perp$ components of four vectors. Those are independent of $E$; and so, 
\begin{align}
 H_{E\perp} &=P_{\perp}^2+4p_{\perp}^2+4\left\{\kappa^2{x}_{\perp}^2+(mc)^2\right\}, \\
 T_{E\perp} &=P_\perp\cdot p_\perp.
 \end{align}
Meanwhile, the ($H_{E\|},T_{E\|}$) are $E$ dependent parts of the constraints; that is,
  \begin{align}
 H_{E\|} &=\Pi_{\|}^2+4\left(\pi_\|^2+\kappa^2 x_{\|}^2\right), \\
 T_{E\|} &=\Pi_\|\cdot\pi_{\|}.
 \end{align}
 It is obvious that the presence of $E\neq 0$ does not alter the $\perp$ components of the canonical variables, though the $\|$ components of the canonical variables are affected by the electric field. As a result, the Poisson brackets of those variables become, for $\mu,\nu=0,1$,
 \begin{align}
   \{\Pi_{\|}^\mu, \Pi_{\|}^\nu\} &=\frac{gE}{c}\Delta\tilde{g}\epsilon^{\mu\nu}, \label{PiPi}\\
   \{\pi_{\|}^\mu, \pi_{\|}^\nu\} &=\frac{gE}{c}\frac{\Delta\tilde{g}}{4}\epsilon^{\mu\nu}, \label{pipi}\\
   \{\Pi_{\|}^\mu, \pi_{\|}^\nu\} &=\frac{gE}{c}\epsilon^{\mu\nu}, \label{Pipi}
\end{align}
with $\{\Pi_\|^\mu,x_\|^\nu\}=0,~\{\pi_\|^\mu,x_\|^\nu\}=-\eta_{\|}^{\mu\nu}$. Here, $\Delta\tilde{g}=\tilde{g}_{(1)}+\tilde{g}_{(2)}$ and $\epsilon^{\mu\nu}=-\epsilon^{\nu\mu} \,(\mu,\nu=0,1;\,\epsilon^{01}=1)$. 

Because of those non-vanishing structures of the right-hand sides of equations (\ref{PiPi})-(\ref{Pipi}), it is not easy to construct a set of first class constraints such as the $(H,T^{(+)})$ in the case of $E=0$. Further, the on-mass-shell structure of the bi-local system described by $H_E=0$ is hard to understand in terms of $(\Pi_\|,\pi_\|)$. To overcome this situation,  let us introduce a new set of momenta $(\Pi_{[g]\|},\pi_{g\|})$ with
\begin{align}
 {\Pi}_{[g]\|} &={\Pi}_{g\|}-\frac{\tilde{g}_*gE}{c}\frac{\tilde{g}_*}{w_g}\tilde{x}_{\|}+\frac{\Delta_g}{w_g}{\pi}_{g\|}, \label{Pi[g]}
\end{align}
to which one can verify
\begin{align}
 \{\Pi_{[g]\|}^\mu,\pi_{g\|}^\nu\} &=\{\Pi_{[g]\|}^\mu,x_\|^\nu\}=0 , \label{Pi[g]-rela} \\
  \{\Pi_{[g]\|}^\mu,\Pi_{[g]\|}^\nu\} &=2\frac{\tilde{g}_*gE}{c}\frac{\tilde{g}_*\Delta_g}{w_g}\epsilon^{\mu\nu}, \label{Pi[g]-Pi[g]}
\end{align}
in addition to  $\{\pi_{g\|}^\mu,\pi_{g\|}^\nu\}=0$ and $\{\pi_{g\|}^\mu,x_\|^\nu\}=-w_g\eta_{\|}^{\mu\nu}$. The equations (\ref{Pi[g]-rela})-(\ref{Pi[g]-Pi[g]}) mean that the $\Pi_{[g]\|}$ is independent of the relative variables, though there remains the effect of $E\neq 0$ for $\Delta_g\neq 0$

Next, rewriting $(H_{E\|},T_{E\|})$ in terms of  $(\Pi_{[g]\|},\pi_{g\|})$, we obtain
\begin{align}
 H_{E\|} &=\frac{1}{\tilde{g}_*}{\Pi}_{[g]\|}^2+\frac{4}{w_g^2}{\pi}_{g\|}^2+\left\{4\kappa^2-\frac{\tilde{g}_*}{w_g^2}\left(\frac{gE}{c}\tilde{g}_*\right)^2\right\}{x}_{\|}^2 \nonumber \\
 &~~~+\frac{2\Delta_g}{w_g}{T}_{E\|}+\frac{2\tilde{g}_*}{w_g}\left(\frac{gE}{c}\right){\Pi}_{[g]\|}\cdot\tilde{x}_{\|}, \\
 T_{E\|} &=\left(\frac{w_g}{\tilde{g}_*}\Pi_{[g]\|}+\frac{gE}{c}\tilde{g}_*\tilde{x}_{\|}\right) \nonumber \\
 &~~~~ \cdot\left\{\frac{1}{w_g}\pi_{g\|}+\frac{\Delta_g}{4w_g}\left(\frac{w_g}{\tilde{g}_*}\Pi_{[g]\|}+\frac{gE}{c}\tilde{g}_*\tilde{x}_{\|}\right)\right\}.
\end{align}
Using these expressions to the functions of primary constraints $(H_{E\|},H_{E\perp};T_{E\|},T_{E\perp})$, one can verify the following algebras:
\begin{align}
\{T_{E\|},H_E\} &=-8\kappa^2\frac{w_g}{\tilde{g}_*}\Pi_{[g]\|}\cdot x_{\|}, \\
 \{T_{E\perp},H_E\} &=-8\kappa^2P_\perp\cdot x_\perp, 
 \end{align}
 from which follows
\begin{align}
  \{T_{E},H_E\}=-8\kappa^2 T_x , \label{TE-dot}
\end{align}
where
\begin{align}
 T_x=\frac{w_g}{\tilde{g}_*}\Pi_{[g]\|}\cdot x_{\|}+P_\perp\cdot x_\perp.
\end{align}
The Eq.(\ref{TE-dot}) says that the time development of $T_E$ by the Hamiltonian $H_E$ produces secondarily the constraint $T_{x}\approx 0$, which does not form a set of the first class constraints with $T_E\approx 0$. To handle this problem, we deal with these constraints so that we first eliminate $T_E$ in $H_E$ by replacing $T_{E\|}=T_E-T_{E\perp}\rightarrow -T_{E\perp}$, and next require an additional constraint $\Pi_{[g]\|}\cdot\tilde{x}_\|\approx 0$. In other words, we use  $H_{E\|}$ in the modified form

\begin{align}
 \mathcal{H}_{E\|} &=\frac{1}{\tilde{g}_*}{\Pi}_{[g]\|}^2+\frac{4}{w_g^2}{\pi}_{g\|}^2+\left\{4\kappa^2-\frac{\tilde{g}_*}{w_g^2}\left(\frac{gE}{c}\tilde{g}_*\right)^2\right\}{x}_{\|}^2 \nonumber \\
 &~~~-\frac{2\Delta_g}{w_g}{T}_{E\perp}
\end{align}
; then, a little calculation leads to
\begin{align}
 \mathcal{H}_E &=\mathcal{H}_{E\|}+\mathcal{H}_{E\perp}~~(\mathcal{H}_{E\perp}\equiv H_{E\perp}) \nonumber \\
  &=\frac{1}{\tilde{g}_*}\Pi_{[g]\|}^2+\frac{4}{w_g^2}\pi_{g\|}^2+\left\{4\kappa^2-\frac{\tilde{g}_*}{w_g^2}\left(\frac{gE}{c}\tilde{g}_*\right)^2\right\}{x}_{\|}^2 \nonumber\\
 &+\frac{\tilde{g}_*}{w_g^2}P_{\perp}^2+4\left\{p_{\Delta\perp}^2+\kappa^2{x}_{\perp}^2+(mc)^2 \right\},
\end{align}
where
\begin{align}
 p_{\Delta\perp}=p_{\perp}-\frac{\Delta_g}{4w_g}P_{\perp}.
\end{align}
Using this expression of $\mathcal{H}_E$,  the time development of $\Pi_{[g]\|}\cdot\tilde{x}_\|$  generates the constraints $\big({\Pi}_{[g]\|}\cdot{x}_{\|},\,{\Pi}_{[g]\|}\cdot{\pi}_{\|},\,{\Pi}_{[g]\|}\cdot\tilde{\pi}_{\|}\big)\approx 0$ secondarily according to the algebras 
\begin{align}
 \{{\Pi}_{[g]\|}\cdot\tilde{x}_{\|},\mathcal{H}_{E}\} &=-4\frac{gE}{c}\frac{\Delta_g}{w_g}\tilde{g}_*{\Pi}_{[g]\|}\cdot{x}_{\|} \nonumber\\
 &~~~~+8\frac{1}{w_g}\Pi_{[g]\|}\cdot\tilde{\pi}_{g\|}, \label{secondary-i}\\
  \{{\Pi}_{[g]\|}\cdot{x}_{\|},\mathcal{H}_{E}\} &=-4\frac{gE}{c}\frac{\Delta_g}{w_g}\tilde{g}_*{\Pi}_{[g]\|}\cdot\tilde{x}_{\|} \nonumber \\
  &~~~~ +8\frac{1}{w_g}{\Pi}_{[g]\|}\cdot{\pi}_{g\|} ,
\end{align}
\begin{align}
   \{{\Pi}_{[g]\|}&\cdot\tilde{\pi}_{g\|},\mathcal{H}_{E}\}=-4\frac{gE}{c}\frac{\Delta_g}{w_g}\tilde{g}_*{\Pi}_{[g]\|}\cdot{\pi}_{g\|} \nonumber  \\
   &-2 w_g\left\{4\kappa^2-\frac{\tilde{g}_*}{w_g^2}\left(\frac{gE}{c}\tilde{g}_*\right)^2\right\}{\Pi}_{[g]\|}\cdot\tilde{x}_{\|},  \\
 \{{\Pi}_{[g]\|}&\cdot{\pi}_{g\|},\mathcal{H}_{E}\}=-4\frac{gE}{c}\frac{\Delta_g}{w_g}\tilde{g}_*{\Pi}_{[g]\|}\cdot\tilde{\pi}_{g\|} \nonumber \\
 &-2 w_g\left\{4\kappa^2-\frac{\tilde{g}_*}{w_g^2}\left(\frac{gE}{c}\tilde{g}_*\right)^2\right\}{\Pi}_{[g]\|}\cdot{x}_{\|}. \label{secondary-f}
 \end{align}
It should be noticed that $T_{x\|}=\frac{w_g}{\tilde{g}_*}{\Pi}_{[g]\|}\cdot{x}_{\|}\approx 0$ arises as a crew of those secondary constraints; and so, we need to require $T_{x\perp}=P_\perp\cdot x_\perp\approx 0$ separately.
 
 The equations (\ref{secondary-i})-(\ref{secondary-f}) also mean that a linear combination of ${\Pi}_{[g]\|}\cdot\tilde{x}_{\|}\sim {\Pi}_{[g]\|}\cdot{x}_{\|}$ is able to form a set of first class constraint with $\mathcal{H}_E$. That is, under an appropriate choice of $a_i\,(i=1\sim4)$, the combination
\begin{align}
 \Lambda_{\|} &=a_1{\Pi}_{[g]\|}\cdot\tilde{x}_{\|}+a_2{\Pi}_{[g]\|}\cdot{x}_{\|}
  \nonumber \\
 &~~~+a_3{\Pi}_{[g]\|}\cdot\tilde{\pi}_{g\|}+a_4{\Pi}_{[g]\|}\cdot{\pi}_{g\|}
 \end{align}
will satisfy the closed algebra $\{\Lambda_{\|},\mathcal{H}_E\}=\bar{k}\Lambda_{\|}$. Then the $\Lambda_\|\approx 0$ will play a role of the physical state condition which should be compared with $\hat{T}^{(\pm)}|\Phi\rangle=0$ in the free bi-local fields; we must discuss these points in detail.
 
\section{Physical state conditions}
 
In the quantized theories, the Poisson brackets must be replaced by the commutation relations according to $[\hat{f},\hat{g}]=i\hbar\{f,g\}$. Then, according to Eqs.(\ref{secondary-i})-(\ref{secondary-f}), the algebra $[\hat{\Lambda}_\|,\hat{\mathcal{H}}_E]=k\hat{\Lambda}_\|$ gives rise to the following system of simultaneous equations:
 \begin{align}
 ka_1 &=i\hbar\left[-4\frac{gE}{c}\frac{\Delta_g}{w_g}\tilde{g}_*\times a_2 \right. \nonumber\\
  &~~\left. -2 w_g\left\{4\kappa^2-\frac{\tilde{g}_*}{w_g^2}\left(\frac{gE}{c}\tilde{g}_*\right)^2\right\}\times a_3 \right], \label{sys-eqs-1}\\
 ka_2 &=i\hbar\left[-4\frac{gE}{c}\frac{\Delta_g}{w_g}\tilde{g}_*\times a_1 \right. \nonumber \\
  & \left. -2w_g\left\{4\kappa^2-\frac{\tilde{g}_*}{w_g^2}\left(\frac{gE}{c}\tilde{g}_*\right)^2\right\}\times a_4 \right], \\
 ka_3 &=i\hbar\left[8\frac{1}{w_g}\times a_1-4\frac{gE}{c}\frac{\Delta_g}{w_g}\tilde{g}_*\times a_4 \right], \\
 ka_4 &=i\hbar\left[ 8\frac{1}{w_g}\times a_2-4\frac{gE}{c}\frac{\Delta_g}{w_g}\tilde{g}_*\times a_3 \right]. \label{sys-eqs-4}
 \end{align}
 
To write down the solutions for $(a_i)$, it is convenient to use the notations
\begin{align}
 \bm{u}=\begin{bmatrix}a_1 \\ a_2\end{bmatrix},~~\bm{v}=\begin{bmatrix}a_3 \\ a_4\end{bmatrix},
 ~~\bm{n}_{\pm}=\begin{bmatrix}1 \\ \pm 1 \end{bmatrix},
\end{align}
in terms of which one can write
\begin{align}
 \hat{\Lambda}_{\|} &=\left(\hat{\Pi}_{[g]\|}\cdot\tilde{x}_{\|},\hat{\Pi}_{[g]\|}\cdot{x}_{\|}\right)\bm{u} \nonumber \\
  &~~+\left(\hat{\Pi}_{[g]\|}\cdot\tilde{\hat{\pi}}_{[g]\|},\hat{\Pi}_{[g]\|}\cdot{\hat{\pi}}_{\|}\right)\bm{v}. \label{Lambda-||}
\end{align}
Then the solutions can be represented as (Appendix B)
\begin{align}
\bm{u}^{(\pm)}=N\bm{n}_{\pm}, ~~\bm{v}^{(\pm)}_{\epsilon}=\epsilon N\frac{i}{w_gK_g}\bm{n}_{\pm}~~(\, \epsilon=\pm \,), \label{uvn+-} 
\end{align}
where $N$ is a normalization constant. The $K_g$ is the following function of $E$, which  has a meaning of an effective coupling constant for $\|$ components of the bi-local systems, with the step limits:
\begin{align}
 K_g &=\sqrt{\kappa^2-\frac{\tilde{g}_*}{w_g^2}\left(\frac{\tilde{g}_*gE}{2c}\right)^2}  \label{Kg} \\
 &\stackrel{|g_{(i)}|=g}{\longrightarrow} \kappa_g=\sqrt{\kappa^2-\left(\frac{gE}{2c}\right)^2}~\stackrel{g=0}{\rightarrow}~ \kappa \,.
 \end{align}

Substituting $(\bm{u}^{(\pm)},\bm{v}^{(\pm)}_\epsilon)$ for Eq.(\ref{Lambda-||}), we obtain
\begin{align}
   \hat{\Lambda}^{(\pm)}_{\epsilon\|} &=N\left\{\left(\hat{\Pi}_{[g]\|}\cdot\tilde{x}_{\|} \pm \hat{\Pi}_{[g]\|}\cdot{x}_{\|}\right) \right. \nonumber \\
   &~~~\left. +\epsilon \frac{i}{w_gK_g}\left(\hat{\Pi}_{[g]\|}\cdot\tilde{\hat{\pi}}_{[g]\|} \pm \hat{\Pi}_{[g]\|}\cdot{\hat{\pi}}_{\|}\right)\right\} \nonumber \\
 &=N\left\{ \hat{\Pi}_{[g]\|}\cdot\left(\tilde{x}_{\|} +\epsilon \frac{i}{w_gK_g}\tilde{\hat{\pi}}_{\|}\right) \right. \nonumber\\
 & ~~~~~~~ \left. \pm \hat{\Pi}_{[g]\|}\cdot\left({x}_{\|} +\epsilon \frac{i}{w_gK_g}{\hat{\pi}}_{\|}\right) \right\},
\end{align}
where the double sign $\pm$ corresponds to that of $\bm{n}_{\pm}$.  For reference, the $k$ associated with $\hat{\Lambda}^{(\pm)}_{\epsilon\|}$ is given in Eq.(\ref{k+-}).

The next step is to introduce the ladder operators $(\hat{a}^\mu_{g\|},\hat{a}^{\dag \nu}_{g\|}),\,(\mu=0,1)$ which play the role of the oscillator variables for the $\mathcal{H}_{E\|}$ and the $\hat{\Lambda}^{(\pm)}_{\epsilon\|}$. Remembering $[\hat{\pi}_{g\|}^\mu,x_{\|}^\nu]=-i\hbar w_g\eta^{\mu\nu}_{\|}$ and $[\hat{\pi}_{g\|}^\mu, \hat{\pi}_{g\|}^\nu]=0$, this can be done straightforward; and, we obtain
\begin{align}
  \hat{a}_{g\|}^\mu &=\sqrt{\frac{K_gw_g}{2\hbar}}\left(x_{\|}^\mu+\frac{i}{w_gK_g}\hat{\pi}_{g\|}^\mu \right) , \label{a-g} \\
  \hat{a}_{g\|}^{\dag\mu} &=\sqrt{\frac{K_gw_g}{2\hbar}}\left(x_{\|}^\mu-\frac{i}{w_gK_g}\hat{\pi}_{g\|}^\mu \right) . \label{a-g-2}
\end{align}
Then one can verify $[\hat{a}^\mu_{g\|},\hat{a}^{\dag^\nu}_{g\|}]=\eta_{\|}^{\mu\nu}=\eta^{\mu\nu},\,(\mu,\nu=0,1)$ in addition to the step limits
\begin{align}
 \hat{a}_{g\|}^\mu &\stackrel{|g_{(i)}|\rightarrow g}{\longrightarrow}~ \hat{a}^\mu_{*\|}=\frac{\kappa_g}{2\hbar}\sqrt{\frac{1}{\hbar}}\left(x_{\|}^\mu+\frac{i}{\kappa_g}\hat{\pi}_{g\|}^\mu \right) \\
 &~~\stackrel{g\rightarrow 0~}{\longrightarrow}~~~ \hat{a}^\mu_{\|}=\sqrt{\frac{\kappa}{2\hbar}}x_{\|}^\mu+\frac{i}{\sqrt{2\hbar\kappa}}\hat{p}^\mu_{\|}. 
\end{align}
In terms of those ladder operators, with the setting $N=\sqrt{\frac{K_gw_g}{2\hbar}}$, we obtain the expressions
\begin{align}
\begin{split}
 \hat{\Lambda}^{(+)}_{+\|} &=\hat{\Pi}_{[g]\|}\cdot\tilde{\hat{a}}_{g\|}+\hat{\Pi}_{[g]\|}\cdot\hat{a}_{g\|}=2\hat{\Pi}_{[g]\|}^{-}\hat{a}^{+}_g, \\
 \hat{\Lambda}^{(+)}_{-\|} &=\left(\hat{\Lambda}^{(+)}_{+\|}\right)^\dag=2\hat{\Pi}_{[g]\|}^{-}\hat{a}^{+\dag}_g.
 \end{split} \label{a+}
\end{align} 
Here,
\begin{align}
 \hat{\Pi}_{[g]\|}^{\pm} &=\frac{1}{\sqrt{2}}\left(\pm \hat{\Pi}_{[g]}^{0}+\hat{\Pi}_{[g]}^{1}\right), \\
 \hat{a}_{g}^{\pm} &=\frac{1}{\sqrt{2}}\left( \pm\hat{a}_{g}^{0}+\hat{a}_{g}^{1}\right),
 \end{align}
and $(\hat{a}^{\pm}_g,\hat{a}^{\pm\dag}_g)$ are the ladder operators characterized by the lightlike commutation relations
\begin{align}
\begin{split}
 [\hat{a}_g^{+},\hat{a}_g^{-\dag}] &= [\hat{a}_g^{-},\hat{a}_g^{+\dag}]=1, \\
 [\hat{a}_g^{+},\hat{a}_g^{+\dag}] &= [\hat{a}_g^{-},\hat{a}_g^{-\dag}]=0.
 \end{split} \label{pm-commutators}
\end{align}
Similarly we get the expressions
\begin{align}
\begin{split}
 \hat{\Lambda}^{(-)}_{+\|} &=\hat{\Pi}_{[g]\|}\cdot\tilde{\hat{a}}_{g\|}-\hat{\Pi}_{[g]\|}\cdot\hat{a}_{g\|}=-2\hat{\Pi}^{+}_{[g]\|}\hat{a}^{-}_g, \\
 \hat{\Lambda}^{(-)}_{-\|} &=\left(\hat{\Lambda}^{(-)}_{+\|}\right)^\dag=-2\hat{\Pi}^{+}_{[g]\|}\hat{a}^{-\dag}_g.
 \end{split} \label{a_}
 \end{align}
 The equations  (\ref{a+}) and (\ref{a_}) imply that $\hat{\Lambda}^{(*)}_{+\|}$ and $\hat{\Lambda}^{(*)}_{-\|}$ are respectively the counterparts of $\hat{T}^{(+)}$ and $\hat{T}^{(-)}$ in the free bi-local fields. Therefore, as for the $\|$ components of dynamical variables, the physical state conditions in the presence of a constant electric field should be
\begin{align}
 \hat{\Pi}^{+}_{[g]\|}\hat{a}_g^{-}|\Phi_{\rm phy}\rangle=0~~\mbox{or}~\hat{\Pi}^{-}_{[g]\|}\hat{a}_g^{+}|\Phi_{\rm phy}\rangle=0\,. \label{a+-}
 \end{align}
These two kinds of supplementary conditions are standing on the same footing as the physical state conditions. Thus, the physical states are characterized by the on-mass shell equation $\hat{\mathcal{H}}_E|\Phi_{\rm phy}\rangle=0\,(\hat{\mathcal{H}}_E=\hat{\mathcal{H}}_{E\|}+\hat{\mathcal{H}}_{E\perp})$ in addition to one of supplementary conditions in Eq.(\ref{a+-}). 

To write down the operators $(\hat{\mathcal{H}}_{E\|},\hat{\mathcal{H}}_{E\perp})$ explicitly, it is useful to introduce the ladder operators for $\perp$ variables in addition to those of Eqs.(\ref{a-g}) and (\ref{a-g-2}) so that
\begin{align}
 \hat{a}_{g\perp}^i &=\frac{1}{\sqrt{2\hbar\kappa}}\left(\kappa{x}_{\perp}^i+i\hat{p}_{\Delta\perp}^i\right), \label{a-perp} \\
 \hat{a}_{g\perp}^{i \dag} &=\frac{1}{\sqrt{2\hbar\kappa}}\left(\kappa{x}_{\perp}^i-i\hat{p}_{\Delta\perp}^i\right),
\end{align} 
$(i=2,3)$, to which one can verify $\left[\hat{a}_{g\perp}^i,\hat{a}_{g\perp}^{j\dag}\right]=\eta_{\perp}^{ij}=\delta^{ij}$ and such a limit as $\hat{a}_{g\perp}^i\rightarrow \hat{a}_{\perp}^i     
=\frac{1}{\sqrt{2\hbar\kappa}}(\kappa x_\perp^i+i\hat{p}_\perp^i)$ according as $\Delta_g\rightarrow 0$; here, $(\hat{a}_\perp,\hat{a}_\perp^\dag)$ are the ladder operators for a free bi-local field. In terms of those ladder operators, we can write
\begin{align}
 \hat{\mathcal{H}}_{E\|} &=\frac{1}{\tilde{g}_*}\hat{\Pi}_{[g]\|}^2+\frac{1}{2}\left(8\hbar\frac{K_g}{w_g}\right)\left\{\hat{a}_{g\|}^{\dag\mu},\hat{a}_{g\|\mu} \right\} \nonumber \\
 &=\frac{1}{\tilde{g}_*}\hat{\Pi}_{[g]\|}^2+\left(8\hbar\frac{K_g}{w_g}\right)\left\{\sum_{\pm}\hat{a}_g^{\pm\dag}\hat{a}_g^{\mp}+1\right\}, \\
 \hat{\mathcal{H}}_{E\perp} &=\frac{\tilde{g}_*}{w_g^2}\hat{P}_{\perp}^2+4\hbar\kappa\left\{\hat{a}_{g\perp}^{\dag\mu},\hat{a}_{g\perp\mu}\right\}+4(mc)^2 \nonumber \\
 &=\frac{\tilde{g}_*}{w_g^2}\hat{P}_{\perp}^2+8\hbar\kappa\left(\hat{a}_{g\perp}^{\dag}\cdot\hat{a}_{g\perp}+1\right)+4(mc)^2,
\end{align}
where we have used $\hat{a}^\dag_{g\|}\cdot\hat{a}_{g\|}=\hat{a}_{g}^{+\dag}\hat{a}_{g}^{-}+\hat{a}_{g}^{-\dag}\hat{a}_{g}^{+}$.

\section{Structure of the physical states}

In what follows, without loss of generality, we adopt the $\hat{a}^{-}$ type of condition in Eq.(\ref{a+-}) to characterize the physical states. Then, the master wave equation for the bi-local fields should be set
\begin{align}
\hat{\mathcal{H}}_E &|\Phi_{\rm phy}\rangle =0, \label{MW-eq} \\
\mbox{where}\hspace{15mm} &  \hspace{7cm} \nonumber \\
 \hat{\mathcal{H}}_E =\Bigg(\frac{1}{\tilde{g}_*}\hat{\Pi}_{[g]\|}^2 &+\frac{\tilde{g}_*}{w_g^2}\hat{P}_{\perp}^2\Bigg) +\left(8\hbar\frac{K_g}{w_g}\right)\hat{a}_g^{-\dag}\hat{a}_g^{+}  \nonumber \\
 &+8\hbar\kappa\hat{a}_{g\perp}^{\dag}\cdot\hat{a}_{g\perp}+(m_gc)^2, \label{Wave operator}
 \end{align}
and the effective rest mass $m_g$ is given by
 \begin{align}
 (m_gc)^2=4\left(2\hbar\frac{K_g}{w_g}+2\hbar\kappa+(mc)^2\right) . \label{effective mass}
\end{align}
In addition to Eq.(\ref{MW-eq}), the state $|\Phi_{\rm phy}\rangle$ must satisfy 
 \begin{align}
 \hat{\Pi}_{[g]\|}^{+}\hat{a}^{-}_g|\Phi_{\rm phy}\rangle=0. \label{supple-condition}
\end{align}

Since each terms of $\hat{\Pi}_{[g]\|}^2$, $\hat{P}^2_\perp$, $\hat{a}^{-\dag}\hat{a}^{+}$ and $\hat{a}^{\dag}_{g\perp}\cdot\hat{a}_{g\perp}$ in $\hat{\mathcal{H}}_E$ are commutable with others, those terms have simultaneous eigenstates. In the beginning, as for the operators $\hat{N}^{\pm}_g=\hat{a}_g^{\pm\dag}\hat{a}_g^{\mp}$, the eigenvalue equation $\hat{N}_g^{\pm}|(n)_{\pm}\rangle=n|(n)_{\pm}\rangle$ can be solved in the form
\begin{align}
 |(n)_\pm\rangle &=\frac{1}{\sqrt{n!}}\left(\hat{a}_g^{\pm\dag}\right)^{n}|0_g^{\|}\rangle ~\,(n=0,1,2,\cdots), \label{(n)pm}
\end{align}
where $\hat{a}^{\pm}_{g}|0_g^{\|}\rangle=0$ with $\langle 0_g^{\|}|0_g^{\|}\rangle=1.$ The inner product between those states can be found as $\langle(n)_\pm|(n^{\prime})_\pm\rangle=0 \,\,(n,n^\prime>0)$ and $\langle (n)_\pm|(n^{\prime})_\mp \rangle=\delta_{n,n^{\prime}}$; and so, $\{|(n)_\pm\rangle\,(n\neq 0)\}$ are zero norm states. Further, the supplementary condition (\ref{supple-condition}) removes one of those bases\footnote{
The pair states $\{|(n)_{+}\rangle,|(n)_{-}\rangle\}$ produce the negative norm state $\frac{1}{\sqrt{2}}\left(|(n)_{+}\rangle-|(n)_{-}\rangle\right)$ unless one of pair states is removed by  Eq.(\ref{supple-condition}).
}, the $\{|(n)_{-}\rangle\}$ in this case. Thus, only the state $|(0)_{+}\rangle=|0_g^{\|}\rangle$ is relevant to physical observation.

Secondaly, the operator $\hat{N}_g^{\perp}=\hat{a}^\dag_{g\perp}\cdot\hat{a}_{g\perp}$ has the eigenstates belonging to an eigenvalue $J\,(=0,1,2,\cdots)$ such that
\begin{align}
 |(J,M)_\perp \rangle &=\frac{1}{\sqrt{n_2!n_3!}}\left(\hat{a}_g^{2\dag}\right)^{J+M}\left(\hat{a}_g^{3\dag}\right)^{J-M}|0_g^{\perp}\rangle , \label{n-perp}
\end{align}
where $\hat{a}_g^{i}|0_g^{\perp}\rangle=0\,(i=2,3; \langle 0_g^{\perp}|0_g^{\perp}\rangle=1)$. The normalization $\langle J,M|J^\prime, M^\prime\rangle=\delta_{J,J^\prime}\delta_{M,M^\prime}$ is obvious; and, as the eigenstates of $\hat{N}_{g}^\perp$, $|J,M\rangle$ with $M=J,\cdots,-J$, are degenerate for each $J$. Since the $J$ has the meaning of spin eigenvalue of the bi-local system generated by the rotation around the $\bm{e}^{1}$ axis, the result is in the right.

Thirdly, let us consider the eigenvalues of $\hat{\Pi}^2_{[g]\|}$, which reveal different phases depending on $\Delta_g=0$ or $\Delta_g\neq 0$.  In the case of $\Delta_g=0$, the $\hat{\Pi}^\mu_{[g]\|}\, (\mu=0,1)$ become commutable operators, which are canonically equivalent to $\hat{P}^\mu \,(\mu=0,1)$ (Appendix A). Thus, the $\hat{\Pi}^2_{[g]\|}$ has continuous eigenvalues $P^{\|2}=-(P^0)^2+(P^1)^2\,(P^\mu\in \mathbb{R})$.

On the other hand for $\Delta_g\neq 0$, one can introduce  a canonical pair $\left(\hat{\Pi}^1_{[g]\|},\frac{c}{2gE}\frac{w_g}{\Delta_g\tilde{g}_*^2}\hat{\Pi}^0_{[g]\|}\right)=(\hat{\mathcal{P}},\hat{\mathcal{X}})$ satisfying $[\hat{\mathcal{X}},\hat{\mathcal{P}}]=i\hbar$. Then one can write
 \begin{align}
 \frac{1}{\tilde{g}_*}\hat{\Pi}^2_{[g]\|}=\frac{1}{\tilde{g}_*}\left\{\hat{\mathcal{P}}^2-\left(\frac{2gE}{c}\frac{\Delta_g\tilde{g}_*^2}{w_g}\right)^2\hat{\mathcal{X}}^2\right\}. \label{RHO}
\end{align}
The right-hand side of Eq.(\ref{RHO}) is the Hamiltonian of a repulsive harmonic oscillator\footnote{The inverted harmonic oscillator in other words.}. A convenient way to handle this Hamiltonian is to introduce the ladder operators defined by
\begin{align}
 A &=A^\dag=\sqrt{\frac{\mu\omega}{2\hbar}}\hat{\mathcal{X}}-\frac{1}{\sqrt{2\mu\hbar\omega}}\hat{\mathcal{P}} \label{ladder-A}, \\
 \bar{A} &=\bar{A}^\dag=\sqrt{\frac{\mu\omega}{2\hbar}}\hat{\mathcal{X}}+\frac{1}{\sqrt{2\mu\hbar\omega}}\hat{\mathcal{P}}, \label{ladder-barA} \\
 \omega &=\frac{2gE}{\mu c}\frac{\Delta_g\tilde{g}_*^2}{w_g},
\end{align}
where $\mu$ is a parameter with the dimension of mass. Then by taking $[A,\bar{A}]=i$ into account, one can factorize the right-hand side of Eq.(\ref{RHO}) so that
\begin{align}
  \hat{H}_{\Pi\|}\equiv \frac{1}{\tilde{g}_*}\hat{\Pi}^2_{[g]\|}=\frac{2\mu\omega}{\tilde{g}_*}\left\{-i\hbar\left(\Lambda+\frac{1}{2}\right)\right\} \label{H-||}
\end{align}
where $\Lambda=-i\bar{A}A$ and $\bar{\Lambda}=-iA\bar{A}\,(=\Lambda+1)$.

With the aid of the algebra of those ladder operators, one can show\cite{RHO-PRD} that the states of pairs
\begin{align}
 |\phi_{(n)}\rangle &=\bar{A}^n|\phi_{(0)}\rangle~~( A|\phi_{(0)}\rangle=0), \label{A-base} \\
 |\bar{\phi}_{(n)}\rangle &=A^n|\phi_{(0)}\rangle~~( \bar{A}|\bar{\phi}_{(0)}\rangle=0) \
\end{align}
$(n=0,1,2,\cdots)$ are solutions of eigenvalue equations $\Lambda|\phi_{(n)}\rangle=n|\phi_{(n)}\rangle$ and $\bar{\Lambda}|\bar{\phi}_{(n)}\rangle=-n|\bar{\phi}_{(n)}\rangle$. Further, those states satisfy the normalization
\begin{align}
 \langle\bar{\phi}_{(m)}|\phi_{(n)}\rangle=\delta_{m,n}N_n~~\left(N_n=i^nn!\sqrt{\frac{i}{2\pi}}\right)
\end{align}
in addition to the completeness 
\begin{align}
 1=\sum_{n=0}^\infty \frac{1}{N_n}|\phi_{(n)}\rangle\langle\bar{\phi}_{(n)}|. 
 \end{align}
 
Therefore, the independent basis of physical states fall into two classes: in the first for $\Delta_g=0$, the independent bases take the forms
\begin{align}
  \left(\hat{U}_Ee^{\frac{i}{\hbar}P_{\|}\cdot X_{\|}}\right) &\otimes |(0)_{+}\rangle \otimes |(J,M)_\perp \rangle \otimes e^{\frac{i}{\hbar}P_\perp\cdot X_\perp},
\end{align}
where $\hat{U}_E=e^{\frac{i}{\hbar}\frac{gE}{2c}\tilde{x}^{\|}\cdot X^{\|}}$. Then, Eq.(\ref{MW-eq}) requires the mass-shell condition
\begin{align}
  P_{\|}^2+P_{\perp}^2+8\hbar\kappa J +(m_gc)^2=0. \label{on-mass-shell-1}
\end{align}
Secondly, for $\Delta_g \neq 0$, the independent bases become
\begin{align}
  |\phi_{(n)}\rangle\otimes |(0)_{+}\rangle \otimes |(J,M)_\perp \rangle \otimes e^{\frac{i}{\hbar}P_\perp\cdot X_\perp},
\end{align}
to which Eq.(\ref{MW-eq}) requires
\begin{align}
  \frac{-2\mu\omega i\hbar\left(n+\frac{1}{2}\right)}{\tilde{g}_*}+\frac{\tilde{g}_*}{w_g^2}P_{\perp}^2+8\hbar\kappa J +(m_gc)^2=0. \label{on-mass-shell-2}
\end{align}
In the first class with $\Delta_g=0$, the mass-shell condition (\ref{on-mass-shell-1}) is almost the same as one of free bi-local fields except the $E$ dependent mass term $(m_gc)^2$. On the other hand, for $\Delta_g\neq 0$, the mass-shell condition (\ref{on-mass-shell-2}) requires a complex $P_\perp^2$. This implies that in this case, there are no stable free bi-local fields as a nature of accelerated bi-local systems with net nonzero charges. Furthermore, the dimensionality of spacetime also adds a crucial aspect to those charged bi-local systems. If the bi-local systems exist in two-dimensional $\|$ plane, then there are no $\perp$ degrees of freedoms $(P_\perp,J)$ originally in Eq.(\ref{on-mass-shell-2}); and so, Eq.(\ref{on-mass-shell-2}) leads to $\omega=m_g=0$, which is impossible for $gE\neq 0$.

\section{Ground state structure of bi-local fields under a constant electric field}

The ladder operators $(\hat{a}_g^{\pm},\hat{a}_g^{\pm\dag})$ and $(\hat{a}_g^\perp,\hat{a}_g^{\perp\dag})$ have been defined with the coupling constants $K_g$ and $\kappa$  respectively. Here, the $K_g$ tends to 0 according as $E\rightarrow E_c=\frac{2cw_g\kappa}{\tilde{g}_*g\sqrt{\tilde{g}_*}}$. Since the $\perp$ degrees of freedoms disappear for the bi-local systems in the $\|$ plane, we may regard $E_c$ as the critical electric field, which is related to the dissociation of two constituent particles. 

By Eq.(\ref{a-perp}),  the ground state $|0_g^\perp\rangle$ can be solved as $|0_g^\perp\rangle=e^{\frac{i}{\hbar}\frac{\Delta_g}{4w_g}x_\perp\cdot\hat{P}_\perp}|0^\perp\rangle$ with the ground state for free bi-local fields defined by $\hat{a}_\perp|0^\perp\rangle=0\,(\langle 0^\perp|0^\perp\rangle=1)$; and, the $|0_g^\perp\rangle$ does not depend on $E$ any more. Contrarily,  the ground state $|0_g^{\|}\rangle$ is determined including $E$ explicitly as a parameter; and so, the $|0_g^{\|}\rangle$ is a function of $K_g$ too. Then it becomes important to evaluate the transition amplitude between the ground states with different $K_g$'s in order to study the possibility of the dissociation.

To study the ground state $|0_g^{\|}\rangle$ in detail, we take notice of the representation of $\hat{\pi}_{g\|}$ written by $(\hat{P},\hat{p},\tilde{X},\tilde{x})$ in such a way that
\begin{align}
 \hat{\pi}_{g\|} &=\frac{1}{2}\left(\sqrt{|\tilde{g}_{(2)}|}\hat{\pi}_{(1)}-\sqrt{|\tilde{g}_{(1)}|}\hat{\pi}_{(2)}\right)_{\|} \nonumber \\
  &=-\frac{\Delta_g}{4}\hat{P}_{\|}+w_g\hat{p}_{\|}+\frac{\tilde{g}_*gE}{2c}\left(w_g\tilde{X}+\frac{\Delta_g}{4}\tilde{x}\right)_{\|}. \label{pi-g}
\end{align}
Substituting (\ref{pi-g}) for (\ref{a-g}), we obtain
\begin{align}
\hat{a}_{g\|} &=\sqrt{\frac{w_gK_g}{2\hbar}}\left[x_{\|}+\frac{i}{w_gK_g}\left\{-\frac{\Delta_g}{4}\hat{P}_{\|}+w_g\hat{p}_{\|} \right. \right. \nonumber \\
&\left.\left.+\frac{\tilde{g}_*gE}{2c}\left(w_g\tilde{X}+\frac{\Delta_g}{4}\tilde{x}\right)_{\|} \right\}\right],
\end{align}
from which follows
\begin{align}
 \hat{a}_g^{\pm} &=\sqrt{\frac{w_gK_g}{2\hbar}}\Bigg\{ \sqrt{\frac{\hbar}{2\kappa}}\Big(B_{\pm}\hat{a}^{\pm\dag}+\breve{B}_{\pm}\hat{a}^{\pm} \Big) + i\hat{v}^{\pm} \Bigg\}, \label{ag-pm}
\end{align}
where
\begin{align}
 B_{\pm} &=\Big(1-\frac{\kappa}{K_g}\pm\frac{i}{w_gK_g}\frac{\tilde{g}_*gE}{2c}\frac{\Delta_g}{4}\Big) \nonumber \\
 &=(1-a)\pm ib, \\
 \breve{B}_{\pm} &=\Big(1+\frac{\kappa}{K_g}\pm\frac{i}{w_gK_g}\frac{\tilde{g}_*gE}{2c}\frac{\Delta_g}{4}\Big) \nonumber \\
 &=(1+a)  \pm ib. 
 \end{align}
 Here we have used the notations such that
 \begin{align}
 a &=\frac{\kappa}{K_g}~~(1\leq a \leq \infty),\\
 b &=\frac{\Delta_g}{4w_gK_g}\frac{\tilde{g}_*gE}{2c}=\frac{\Delta_g}{4\sqrt{\tilde{g}_*}}\sqrt{a^2-1} 
 \end{align}
 and
 \begin{align}
 \hat{v}^\pm &=\frac{1}{w_gK_g}\left(-\frac{\Delta_g}{4}\hat{P}^\pm \pm\frac{\tilde{g}_*gE}{2c}w_g X^\pm\right) . \label{v-definition}
 \end{align}
One can verify that  $\hat{v}^\pm$ and $(\hat{a}_g^\pm,\hat{a}_g^{\pm\dag})$ are  commutable with each other due to
 \begin{align}
 [\hat{v}^{-},\hat{v}^{+}]=\frac{\hbar}{2\kappa}\breve{B}_{[-}B_{+]}=i\hbar\frac{2ab}{\kappa}. \label{v-pm}
 \end{align}
Eq.(\ref{v-pm}) means that for $b\neq 0$, the eigenstates $\hat{v}^{\pm}|v^{\pm}\rangle=v^{\pm}|v^{\pm}\rangle$ can be normalized so that
\begin{align}
 \langle v_1^{\pm}|v_2^{\pm}\rangle &=\delta(v_1^{\pm}-v_2^{\pm}), \label{normalization-1}\\
 \langle v^{\pm}|v^{\mp}\rangle &=\sqrt{\frac{1}{2\pi\hbar}\frac{2\kappa}{ab}}e^{\pm \frac{i}{\hbar}\frac{2\kappa}{ab}v^{-}v^{+}}.
 \end{align}
We also note that the unitary transformation: \\ $\hat{v}^{\pm}=-\frac{\Delta_g}{4w_gK_g}\hat{U}^{\pm 1}\hat{P}^{\pm}\hat{U}^{\mp 1}$ with  $\hat{U}=e^{\frac{i}{\hbar}\frac{4}{\Delta_g}\frac{\tilde{g}_*gEw_g}{2c}X^{+}X^{-}}$ relates the eigenvalues of $\hat{v}^\pm$ to those of $\hat{P}^\pm$
\footnote{
The $\hat{v}^{\pm}$ are devided into two algebraic classes according as $\Delta_gE\neq0$ or $\Delta_g E=0$. In the case of $\Delta_g\neq 0$ with $E\neq 0$, the $\hat{v}^\pm$ and the $\hat{P}^\pm$ are not unitarily equivalent because of Eq.(\ref{v-definition}) with non-vanishing right-hande side; and, the successive unitary transformations of $\hat{U}$ and $\hat{V}=\exp\left\{\frac{i}{\hbar}\frac{4\Delta_g c}{\tilde{g}_*gEw_g}\hat{P}^{+}\hat{P}^{-}\right\}$ which is singular at $E=0$, lead to $(\hat{U}\hat{V})^\dag(\hat{v}^{+},\hat{v}^{-})(\hat{U}\hat{V})=\left(-\frac{\Delta_g}{4w_gK_g}\hat{P}^{+},-\frac{\tilde{g}_*gE}{K_g}X^{-}\right)$. In other words, $\hat{v}^{+}$ and $\hat{v}^{-}$ are unitarily equivalent respectively to $\hat{P}^{+}$ and $X^{-}$ apart from coefficients. On the other side, in the case of $\Delta_g\neq 0$ with $E=0$, the Eq.(\ref{v-definition}) is simply reduced to $v^{\pm}=-\frac{\Delta_g}{4\kappa}P^{\pm}$.  The same discussion is available by interchanging the role between $\Delta_g$ and $E$. These properties of $\hat{v}^{\pm}$ imply that the $\Delta_gE$ plays a role of order parameter characterizing equivalent classes of $\hat{v}^\pm$.
}. In this case, since $\hat{U}$ is an operator acting on $\pm$ spaces, it is convenient to realize the eigenstates of $\hat{v}^\pm$ on the product bases $|P^{\|}\rangle\!\rangle=|P^{+}\rangle\otimes|P^{-}\rangle$, where  $\hat{P}^\pm|P^\pm\rangle=P^\pm|P^\pm\rangle$ $(\, \langle P_1^\pm|P_2^\pm\rangle=\delta^{(4)}(P_1^\pm-P_2^\pm) \,)$. Then we can define
\begin{align}
  |(v)^{\pm}\rangle\!\rangle  =\hat{U}^{\pm 1}|P^{\|}\rangle\!\rangle \label{(v)+-} ,
\end{align}
with
\begin{align}
 v^\pm  =-\frac{\Delta_g}{4w_gK_g}P^{\pm}=-\frac{a\Delta_g}{4w_g\kappa}P^{\pm},~~~~~~
 \end{align}
which satisfy $\hat{v}^{\pm}|(v)^{\pm}\rangle\!\rangle=v^\pm|(v)^{\pm}\rangle\!\rangle$ and $\langle\!\langle (v_1)^\pm|(v_2)^\pm\rangle\!\rangle=\delta^{(2)}(P_1^{\|}-P_2^{\|})$; further, the both bases tend to $|P^{\|}\rangle\!\rangle$ according as $E\rightarrow 0$. In terms of $|(v)^{\pm}\rangle\!\rangle$, we can write the ground states annihilated by $\hat{a}_g^{\pm}$ so that (Appendix C)
\begin{align}
  \hat{G}(\hat{v}^{-},\hat{v}^{+})|0^{\|}\rangle\otimes |(v)^{+}\rangle\!\rangle , \label{ground-1}
\end{align}
where $|0^{\|}\rangle$ is the ground state for free bi-local systems defined by $\hat{a}^\pm|0^{\|}\rangle=0~(\, \langle 0^{\|}|0^{\|}\rangle=1\,)$, and 
\begin{align}
 \hat{G}(\hat{v}^{-},\hat{v}^{+}) &=e^{-\alpha\hat{a}^{+\dag}\hat{a}^{-\dag}-i\hat{\beta}(\hat{v}^{-})\hat{a}^{+\dag}}e^{-i\hat{\gamma}(\hat{v}^{+})\hat{a}^{-\dag}} 
 \end{align}
 with
 \begin{align}
  \alpha &=\frac{B_{-}}{\breve{B}_{-}},~\hat{\beta}(\hat{v}^{-})=\frac{1}{\breve{B}_{-}}\sqrt{\frac{2\kappa}{\hbar}}\hat{v}^{-},~\hat{\gamma}(\hat{v}^{+})=\frac{1}{\breve{B}_{+}}\sqrt{\frac{2\kappa}{\hbar}}\hat{v}^{+} .
\end{align}
Since it holds that $\hat{a}_g^{\pm}\hat{G}|0^{\|}\rangle=0$, the ground state equation for $|0_g^{\|}\rangle$ is satisfied by $\hat{G}|0^{\|}\rangle$ only; however, the $|(v)^{+}\rangle\!\rangle$ in (\ref{ground-1}) allows us to replace $\hat{v}^{+}$ in $\hat{G}$ simply with the eigenvalue $v^{+}$.

Now, a simple norm of the state (\ref{ground-1}) is divergent, since $|(v)^{+}\rangle\!\rangle$ takes continuous spectrum; however, we can construct the following ground state having a similar normalization to $|P^{\|}\rangle\!\rangle=|P^{+}\rangle\otimes|P^{-}\rangle$:
\begin{align}
 |0^E_g,(v)^{+}\rangle\!\rangle \equiv \hat{N}\big[\hat{G}(\hat{v}^{-},\hat{v}^{+})\big]|0^{\|}\rangle\otimes|(v)^{+}\rangle\!\rangle. \label{modified ground state}
 \end{align}
Here, the $\hat{N}[\hat{G}]$ is the mapping of $\hat{G}$ defined by the weighted the matrix elements
\footnote{
One can also define this mapping operationally so that
\begin{align*}
 \hat{N}\big[\hat{G}\big]=\int dv^{-}\int dv^{+}N_{v^{-},v^{+}}\hat{I}_v^{-}\hat{G}\hat{I}_v^{+} ~~\left(\, \hat{I}_v^{\pm}=|v^{\pm}\rangle\langle v^{\pm}| \, \right).
 \end{align*} 
 }
\begin{align}
 \langle v^{-}|\hat{N}\big[\hat{G}\big]|v^{+}\rangle \equiv N_{v^{-},v^{+}} \langle v^{-}|\hat{G}|v^{+}\rangle
\end{align}
with
\begin{align}
 N_{v^{-},v^{+}} &=N_0e^{-\frac{\kappa}{a\hbar}\frac{(1+a)}{(1+a)^2+b^2}v^{-}v^{+}},  \label{Nv_mp} \\
 N_0 &=\sqrt{\frac{4a}{(1+a)^2+b^2}\left|\frac{(1+a)^2-b^2}{(1+a)^2+b^2}\right|}. \label{N_0}
\end{align}
Then, as shown in Appendix C, it can be verified that
\begin{align}
\langle\!\langle 0^E_g,(v_1)^{+}|0^E_g,(v_2)^{+}\rangle\!\rangle=\delta^{(2)}(P_1^{\|}-P_2^{\|}).
\end{align}
Further, by taking the $v^{-}$ representation characterized by
 $\langle v^{-}|\hat{a}_g^{\pm}=(\hat{a}_g^\pm)_{v^{-}}\langle v^{-}|$, one can find
  \begin{align}
 (\hat{a}_g^\pm)_{v^{-}} &\langle v^{-}|N_{v^{-},v^{+}}[\hat{G}]|0^{\|}\rangle|v^{+}\rangle \nonumber \\
 &=N_{v^{-},v^{+}}\langle v^{-}|\hat{a}_g^\pm\hat{G}|0^{\|}\rangle|v^{+}\rangle=0.
 \end{align} 
  In this sense, the ground state equation is satisfied by $\hat{N}\big[\hat{G}\big]|0^{\|}\rangle$ too. 

Now, by taking $(\alpha,\breve{B}_\pm,a,b)\rightarrow (0,2,1,0)~(E\rightarrow 0)$ into account, the ground state (\ref{modified ground state}) has the limit
\begin{align}
 |\emptyset,{\rm v}\rangle\!\rangle &=\lim_{E\rightarrow 0}|0_g^E,(v)^{+}\rangle\!\rangle =e^{-\frac{\kappa}{2\hbar}\hat{\rm v}^{-}\hat{\rm v}^{+}} \nonumber \\
 &\times e^{-\frac{i}{2}\sqrt{\frac{2\kappa}{\hbar}}\hat{\rm v}^{-}\hat{a}^{+\dag}}e^{-\frac{i}{2}\sqrt{\frac{2\kappa}{\hbar}}\hat{\rm v}^{+}\hat{a}^{-\dag}}|0^{\|}\rangle\otimes |P^{\|}\rangle\!\rangle .  \label{ground-2}
 \end{align}
Here, $\hat{\rm v}^\pm\equiv -\frac{\Delta_g}{4w_g\kappa}\hat{P}^\pm$, to which one can verify $[\hat{\rm v}^{-},\hat{\rm v}^{+}]=0$. Further, in the limit $E\rightarrow 0$, this ground state satisfies
 \begin{align}
 \hat{\rm v}^\pm |\emptyset,{\rm v}\rangle\!\rangle &={\rm v}^\pm|\emptyset,{\rm v}\rangle\!\rangle , \\
 \langle\!\langle \emptyset,{\rm v}_1|\emptyset,{\rm v}_2\rangle\!\rangle &=\delta^{(2)}(P_1^{\|}-P_2^{\|}) .\hspace{10mm}
 \end{align}
 The Eqs.(\ref{modified ground state}) and (\ref{ground-2}) allow us to evaluate the transition between $|\emptyset,v\rangle\!\rangle$, the ground state of free bi-local fields, and $|0_g^E,(v)^{+}\rangle\!\rangle$, a ground state of the bi-local fields under $E\neq 0$. The transition amplitude becomes
\begin{align}
 \langle\!\langle \emptyset,{\rm v}_1|0_g^E,(v)_2^{+}\rangle\!\rangle &=N_0e^{\theta_{12}}D_{12} , \label{transition-amp}
\end{align}
where 
\begin{align}
 D_{12} &=\left(\frac{\Delta_g}{4w_gK_g}\right)^2\left(\frac{1}{2\pi\hbar}\frac{\kappa}{ab}\right) e^{-\frac{i}{\hbar}\frac{\kappa}{ab}v_{12}^{-}v_{12}^{+}} ,\\
 \mbox{Re}(\theta_{12}) &=\frac{\kappa}{\hbar}\frac{(1+a)}{(1+a)^2+b^2}\Bigg[-\frac{(a-1)^2}{a}{\rm v}_1^{-}{\rm v}_1^{+} \nonumber \\
 &+\frac{a-1}{a}({\rm v}_1^{-}v_{21}^{+}-v_{21}^{-}{\rm v}_1^{+})+\frac{1}{a}v_{21}^{-}v_{21}^{+}\Bigg] , \label{Re-theta}\\  
 \mbox{Im}(\theta_{12}) &=\frac{\kappa}{\hbar}\frac{b}{(1+a)^2+b^2}\left[-a{\rm v}_1^{-}{\rm v}_1^{+}- {\rm v}_1^{-}v_2^{+} \right.  \nonumber \\
  &\left. +(2{\rm v}_1^{-}-v_2^{-}){\rm v}_1^{+} \right] ,
 \end{align}
and we have used the notation $f_{ij}=f_i-f_j$. 

The amplitude (\ref{transition-amp}) is consisting of two factors; $N_0e^{\theta_{12}}$ and $D_{12}$. Here, the  $D_{12}$ has the origin in the background momentum eigenstate $|P^{\|}\rangle\!\rangle$ in Eq.(\ref{ground-1}); and it has the limit
\begin{align}
 \lim_{E\rightarrow 0}D_{12}=\left(\frac{\Delta_g}{4w_gK_g}\right)^2\delta^{(2)}(v_{12}^\pm)\Bigg|_{a=1}=\delta^{(2)}(P_{12}^{\pm}). \label{D-limit}
 \end{align}
In this limit, in which $a=1$ and $P_1^\pm=P_2^\pm$, one can easily find $\big|N_0e^{\theta_{12}}\big|^2=1$; this implies that the factor $N_0 e^{\theta_{12}}$ is essential to study the transition amplitude under background momenta $P_i^\pm\,(i=1,2)$ .   

Now, the $\big|N_0e^{\theta_{12}}\big|^2$, the probability density under some normalization, is a function of $a$, which runs from $1$ to $\infty$ according as $E$ increases from $0$ to $E_c=\frac{2cw_g\kappa}{\tilde{g}_*g\sqrt{\tilde{g}_*}}$. To see the explicit form of such a density, let us consider a simple case with ${\rm v}_1^{-}=0$. Then by taking into account ${\rm v}_1^{+}=-\left(\frac{\Delta_g}{4w_g\kappa}\right)P_1^{+}$ and $v_2^\pm=-\left(\frac{a\Delta_g}{4w_g\kappa}\right)P_2^{\pm}$, one can verify from Eqs.(\ref{N_0}) and (\ref{Re-theta}) that 
\begin{align}
 \big|N_0e^{\theta_{12}}\big|^2&=\frac{4a}{(1+a)^2+b^2}\left|\frac{(1+a)^2-b^2}{(1+a)^2+b^2}\right| \nonumber \\
 &\times e^{\frac{2\kappa}{\hbar}\frac{(1+a)a}{(1+a)^2+b^2} \left(\frac{\Delta_g}{4w_g\kappa}\right)^2P_2^{-}P_{21}^{+}}. \label{E to 0}
\end{align}
The structure of $\big|N_0e^{\theta_{12}}\big|^2$ as a function of $a$ is dependent on the sign of $P_2^{-}P_{21}^{+}$; however, roughly speaking, the function decreases gradually from $1$ at $a=1$ to $0$ at $a=\infty$ (FIG.2). The result says that the transition amplitude $\langle\!\langle\emptyset,{\rm v}_1|0_g^{E},(v_2)^{+}\rangle\!\rangle$ comes to be zero at $E=E_c$, which is equivalent to $K_g=0$. The binding force between  the constituent particles of the bi-local system embedded in $(\bm{e}^0,\bm{e}^1)$ spacetime vanishes at $E=E_c$; and, the bi-local system becomes classically dissociated one at $E=E_c$. However, the vanishing amplitude $\big|N_0e^{\theta_{12}}\big|^2$ at  $E=E_c$ implies that the dissociation will not arise in  quantized theories.

\begin{figure}
\center
 \includegraphics[width=5cm]{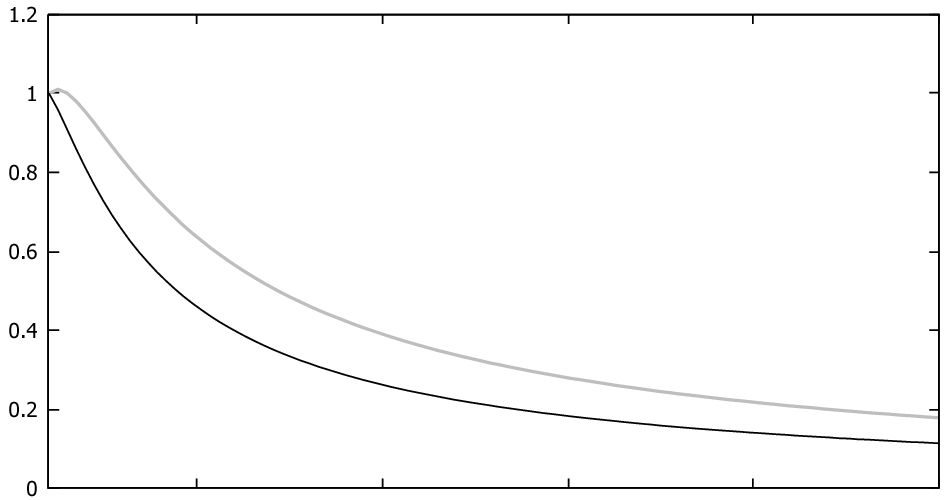}
\caption{The horizontal and the vertical axises designate $a=\frac{\kappa}{K_g}$ and $|N_0e^{\theta_{12}}|^2$ respectively. The gray line is the case of $P_2^{-}P_{21}^{-}>0$, and black line is that of $P_2^{-}P_{21}^{-}<0$.
}
\label{transition}
\end{figure}

\section{The Schwinger effect} 

The action of the bi-local field under a constant electric field is
\begin{align}
 S_{BL}[\Phi_{\rm phy}] &=\langle\Phi_{\rm phy}|\hat{\mathcal{H}}_E|\Phi_{\rm phy}\rangle  \nonumber\\
 &+\lambda\langle\Phi_{\rm phy}|\hat{\Lambda}_{+\|}^{(-)}\hat{\Lambda}_{-\|}^{(-)}|\Phi_{\rm phy}\rangle,
\end{align}
where $\lambda$ is a Lagrange multiplier deriving the physical state condition in Eq.(\ref{supple-condition}). Now, the classical action of gauge field under consideration is $S[A_c]=\int d^4x\left(\frac{E^2}{2}\right)$; and, the one loop correction due to the scalar field $\Phi_{\rm phy}$ adds the quantum effect
\begin{align}
 S_{Q}[A_c]=-i\hbar\ln\left[\left\{\det{}_{\rm phy}\big(\hat{\mathcal{H}}_E\big)\right\}^{-1}\right]
\end{align}
to $S[A_c]$; by taking $\det_{\rm phy}(\hat{\mathcal{H}}_E)=e^{{\rm Tr}_{\rm phy}\ln(\hat{\mathcal{H}}_E)}$ into account, the resultant effective action of gauge field $S_{\rm eff}[A_c]=S[A_c]+S_Q[A_c]$ becomes\cite{RHO-PRD}
\begin{align}
 S_{\rm eff}[A_c]=S[A_c]-i\hbar\int_0^\infty\frac{d\tau}{\tau}e^{-\tau\epsilon}\mbox{Tr}_{\rm phy}\left(e^{-i\tau\hat{\mathcal{H}}_E}\right) \label{S-effective}
\end{align}
disregarding an unimportant addition constant. Here, ${\rm Tr}_{\rm phy}$ is the trace over the physical states, 
the contents of which are depending on $\Delta_g\neq 0$ or $\Delta_g=0$. 

To begin with, we consider the case $\Delta_g\neq 0$, for which $|0_g^{\|}\rangle,\,\{|(J,M)_\perp\rangle\}$ and $\{|\phi_{(n)}\rangle\}$ play the role of physical state basis. The background gauge field $A_c^\mu$, the scalar QED gives rise to the scattering matrix elements $\langle 0_{\rm in}|0_{\rm out}\rangle \sim e^{\frac{i}{\hbar}S_{\rm eff}[A_c]}$ within the one-loop approximation of scalar field $\Phi$; in terms of $S_{\rm eff}[A_c]$, the transition amplitude is written as
\begin{align}
 \big|\langle 0_{\rm in}|0_{\rm out}\rangle\big|^2 \sim e^{-\frac{2}{\hbar}{\rm Im}S_{\rm eff}[A_c]} \label{transition-2}
\end{align}

Now, on the basis of Eq.(\ref{S-effective}),  one can evaluate the Im part of the transition amplitude (\ref{transition-2}) so that 
\begin{align}
 \frac{1}{\hbar}{\rm Im}S_{\rm eff}[A_c] &=-{\rm Re}\int_0^\infty \frac{d\tau}{\tau}e^{-\epsilon\tau}{\rm Tr}_{\rm phy}\left(e^{-i\tau\hat{\mathcal{H}}_E}\right) \\
 &=-{\rm Re}\int_0^\infty \frac{d\tau}{\tau} e^{-i\left\{(m_g c)^2-i\epsilon\right\}\tau} \nonumber \\
 &\times {\rm{Tr}}_\Pi e^{-i\tau\hat{H}_{\Pi\|}}\times {\rm Tr}_{P_\perp}e^{-i\tau\frac{\tilde{g}_*}{w_g^2}\hat{P}_\perp^2} \nonumber \\
 &\times {\rm Tr}_{a\|}e^{-i\tau\hat{H}_{a\|}} \times{\rm Tr}_{a_\perp}e^{-i\tau\hat{H}_{a\perp}}, \label{Im S_eff}
\end{align}
where $\hat{H}_{\Pi\|}$ is the Hamiltonian given in Eq.(\ref{H-||}), and
\begin{align}
 \hat{H}_{a\|} &=\left(\frac{8\hbar K_g}{w_g}\right)\hat{a}_g^{-\dag}\cdot\hat{a}_g^{+}, \\
 \hat{H}_{a\perp} &=8\hbar\kappa\hat{a}_{g\perp}^\dag\cdot\hat{a}_{g\perp} ~.
\end{align}
The physical base relevant to the trace for $ \hat{H}_{a\|}$ is $|(0)_{+}\rangle=|0_g^{\|}\rangle$ only as discussed in Eq.(\ref{(n)pm}), and we may put ${\rm Tr}_{a\|}e^{-i\tau\hat{H}_{a\|}}$ $=1$. Remembering, further, the eigenvalues of  $\hat{a}_{g\perp}^\dag\cdot\hat{a}_{g\perp}$ written in Eq.(\ref{on-mass-shell-2}), we obtain
\begin{align}
 {\rm{Tr}}_\Pi e^{-i\tau\hat{H}_{\Pi\|}} &=\frac{1}{2\sinh\left(\tau\frac{\hbar\mu\omega}{\tilde{g}_*}\right)}, \label{Tr Pi}\\
 {\rm Tr}_{a_\perp}e^{-i\tau\hat{H}_{a\perp}} &=\left\{\frac{e^{i\tau(4\hbar\kappa)}}{2i\sin\left\{\tau(4\hbar\kappa)\right\}} \right\}^2, \\
  {\rm Tr}_{P_\perp}e^{-i\tau\frac{\tilde{g}_*}{w_g^2}\hat{P}_\perp^2} &=\frac{V_\perp}{(2\pi\hbar)^2}\left(\frac{\pi}{i\tau}\right)\frac{w_g^2}{\tilde{g}_*},
\end{align}
where $V_\perp=(2\pi\hbar)^2\delta^{(2)}(P_\perp=0)$ is a cutoff volume of $X_\perp$ space. Substituting these results for Eq.(\ref{Im S_eff}), we have
\begin{align}
 \frac{1}{\hbar}&{\rm Im}S_{\rm eff}[A_c]=-\frac{w_g^2}{\tilde{g}_*\times}{\rm Re}\int_{0}^\infty \frac{d\tau}{\tau} e^{-i\left\{(m_g c)^2-i\epsilon\right\}\tau} \nonumber \\
 &\times \frac{V_\perp}{(2\pi\hbar)^2}\left(\frac{\pi}{i\tau}\right)\times \frac{1}{2\sinh\left(\tau\frac{\hbar\mu\omega}{\tilde{g}_*}\right)}\times \left\{\frac{e^{i\tau(4\hbar\kappa)}}{2i\sin\left\{\tau(4\hbar\kappa)\right\}} \right\}^2 . \label{Im S_eff 2}
\end{align}
Since $(m_gc)^2-(8\hbar\kappa)>0$ by virtue of Eq.(\ref{effective mass}), the exponential of this integral converges in the direction ${\rm Im}\tau\rightarrow -\infty$; and, we regularize the right-hand side of Eq.(\ref{Im S_eff 2}) by regarding the integration range as the $\epsilon\rightarrow 0$ limit of the contour $C^{+}_\epsilon=\{z=\tau e^{-i\epsilon}-i\epsilon;\, 0\leq \tau<\infty\}$ in the complex plane. That is, we put
\begin{align}
  \frac{1}{\hbar}{\rm Im} &S_{\rm eff}[A_c]:=\frac{w_g^2}{\tilde{g}_*}\times\frac{\pi V_{\perp}}{2^3(2\pi\hbar)^2}{\rm Re}\frac{1}{i}\int_{C_\epsilon^{+}}\frac{dz}{z^2} \nonumber \\
  &\times \frac{e^{-iz\left\{(m_g c)^2-(8\hbar\kappa)\right\}}}{\sinh\left(z\frac{\hbar\mu\omega}{\tilde{g}_*}\right)\sin^2\left\{z(4\hbar\kappa)\right\}}\nonumber \\
 &=\frac{\pi V_\perp}{2^3(2\pi\hbar)^2}\frac{1}{2i}\int_{C_\epsilon^{+}-C_\epsilon^{-}} \frac{dz}{z^2} \nonumber \\
 &\times\frac{e^{-iz\left((m_g c)^2-(8\hbar\kappa)\right)}}{\sinh\left(z\frac{\hbar\mu\omega}{\tilde{g}_*}\right)\sin^2\left\{z(4\hbar\kappa)\right\}}. \label{Im S_eff 3}
\end{align}
\begin{figure}
\center
 \includegraphics[width=4cm]{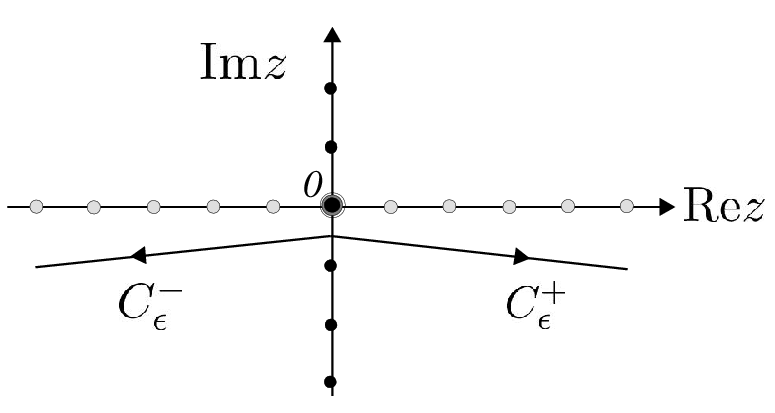}
\caption{The integral in the range $0\leq \tau<\infty$  is dealt as the $\epsilon\rightarrow 0$ limit of the contour integral along $C^{+}_\epsilon$.
 }
\label{Contour}
\end{figure}
Here the $C_\epsilon^{-}$ is the mirror image of $C_\epsilon^{+}$ on the imaginary axis (FIG.{\ref{Contour}). The integration in Eq.(\ref{Im S_eff 3}) can be done by the standard way, and the result is 
\begin{align}
 \frac{1}{\hbar}{\rm Im}S_{\rm eff}[A_c] &=\frac{w_g^2}{\tilde{g}_*}\times\frac{\pi V_\perp}{2^3(2\pi\hbar)^2}\frac{1}{2i}\sum_{n=1}^\infty (-2\pi i) \left\{-\left(\frac{\hbar\mu\omega}{\tilde{g}_*}\right)\right\}\nonumber \\
 &\times \frac{(-1)^n}{(\pi n)^2}\frac{e^{-n\pi\left(\frac{\tilde{g}_*}{\hbar\mu\omega}\right)\left((m_gc)^2-(8\hbar\kappa)\right)}}{-\sinh^2\left\{n\pi\left(\frac{\tilde{g}_*}{\hbar\mu\omega}\right)(4\hbar\kappa)\right\}}.
\end{align}
We may rewrite this formula using the cutoff volume of $X_{\|}$ space given by $V_{\|}\left\{\frac{\sqrt{\hbar\mu\omega}}{(2\pi\hbar)}\right\}^2\sim 1$; then with $\hbar\mu\omega=\frac{2\hbar\tilde{g}_*\Delta_g}{w_g}\frac{\tilde{g}_*gE}{c}$, we finally arrive at the expression
\begin{align}
 \frac{1}{\hbar}{\rm Im}S_{\rm eff}[A_c]=\frac{w_g^2}{\tilde{g}_*}\times\frac{V_{(4)}}{2^3(2\pi\hbar)^4}\left(\frac{2\hbar\Delta_g}{w_g}\frac{\tilde{g}_*gE}{c}\right)^2
 \nonumber \\
 \times\sum_{n=1}^\infty \frac{(-1)^{n+1}}{n^2}\frac{e^{-n\pi\left(\frac{w_g}{2\hbar\Delta_g}\frac{c}{\tilde{g}_*gE}\right)\left\{(m_gc)^2-(8\hbar\kappa)\right\}}}{\sinh^2\left\{n\pi\left(\frac{w_g}{2\hbar\Delta_g}\frac{c}{\tilde{g}_*gE}\right)(4\hbar\kappa)\right\}}, \label{Schwinger eff}
 \end{align}
 where $V_{(4)}=V_{\|}V_\perp$. This result for $\Delta_g\neq 0$ differs from the Schwinger effect for a local scalar field by the factor $\sinh^{-2}(\cdots)$\cite{Schwinger,Berizin-Itykson}. However, if we consider the bi-local fields in the two-dimensional $\|$ plane, the systems do not contain $\perp$ degrees of freedoms. Then the factor $\sinh^{-2}(\cdots)$ in Eq.(\ref{Schwinger eff}) disappear; and, the resultant Schwinger effect has a similar form as one in the scalar QED.
 
Next,  let us consider the case of $\Delta_g=0$ from the beginning, then as discussed in Eq.(\ref{on-mass-shell-1}),  the mass-shell operator (\ref{Wave operator}) is reduced to
\begin{align}
 \hat{\mathcal{H}}_E &\big|_{\Delta_g=0}=\Big(\hat{\Pi}_{[g]\|}^2 +\hat{P}_{\perp}^2\Big)  \nonumber \\
 &+(8\hbar\kappa)\Big(\hat{a}_g^{-\dag}\hat{a}_g^{+}+\hat{a}_{g\perp}^{\dag}\cdot\hat{a}_{g\perp}\Big) +(m_gc)^2
 \end{align}
$\big((m_gc)^2=4(4\hbar\kappa+(mc)^2\big))$; and, the supplementary condition (\ref{supple-condition}) still survive to define physical states . As shown in Eq.(\ref{specific Pi[g]}),  the  $\hat{\Pi}_{[g]\|}^2$ in this case has the same eigenvalues as $\hat{P}_{\|}^2$. Then the  Eq.(\ref{Tr Pi}) is modified so that ${\rm Tr}_{\Pi}e^{-i\tau\hat{H}_{\Pi\|}}=\frac{V_{\|}}{(2\pi\hbar)^2}\left(\frac{\pi}{\tau}\right)$. If we replace the $\sinh$ term in Eq.(\ref{Im S_eff 2}) by $\frac{V_{\|}}{(2\pi\hbar)^2}\left(\frac{\pi}{\tau}\right)$, then the poles in the lower half plane in the integral in Eq.(\ref{Im S_eff 3}) will vanish under the same regularization of $\tau$ integration as that in Eq.(\ref{Im S_eff 3}). This implies that the $\frac{1}{\hbar}{\rm Im}S_{\rm eff}$ comes to be zero; that is, there arise no pair productions of the bi-local fields in this case. Since $\Delta_g=0$ corresponds to neutral bound states, the result is not surprising unless the dissociation of bound sates arises by the electric field.

\section{SUMMARY}

In this paper, we have discussed the formulation of bi-local systems, the classical counterparts of the bi-local fields, under a constant electric field $E\bm{e}^1$ and their physical properties.

The bi-local systems are two particle systems bounded via a relativistic Hooke type of potential; we assign the charges $g_{(1)}\,(>0)$ and $g_{(2)}\,(<0)$ to respective particles. By ordinary, the bi-local systems are formulated as constrained dynamical systems characterized by two first-class constraints: a master wave equation, the mass-shell condition, for the bi-local system and a supplementary condition which can freeze the ghost states caused by relative-time excitation. In many cases, however, the consistency of those constraints is broken by introducing the interaction of the bi-local fields with other fields. In Sec.2 and 3, we have developed a way constructing a set of first class constraints for the bi-local system under a constant electric field. The key was to modify the mass-shell condition by regarding one of secondary constraints as a strong equation.

On this way,  in Sec.4 and 5, we have studied the physical states which can diagonalize  the mass-shell condition of the modified bi-local systems. Then, we had taken notice of that one part of the mass-shell equation is reduced to a Hamiltonian of repulsive harmonic oscillator for $|g_{(1)}|\neq |g_{(2)}|$. The repulsive harmonic oscillator is known to take complex eigenvalues on some basis of eigenstates\cite{RHO-PRD}. So, we had discussed the stability of physical states in the case of $|g_{(1)}|=|g_{(2)}|$ and $|g_{(1)}|\neq |g_{(2)}|$.

Further, in Sec.6, we focused our attention on the properties of ground states, on which the physical states are constructed. In quantum mechanics, the excited states of the bi-local fields are represented by the action of ladder operators $[\hat{a}_g^{\pm},\hat{a}_g^{\mp\dag}]=1$ and $[\hat{a}^i,\hat{a}^{j\dag}]=1$ $,(i,j=2,3)$ on the ground states satisfying $\hat{a}_g^\pm|0_g^{\|}\rangle=\hat{a}^i|0_g^\perp\rangle=0$. In those ground sates, the effect of $E$ appears in $|0_g^{\|}\rangle$ only. In view of this fact, we studied the transition amplitude between the ground states $\{|0_g^{\|}\rangle\}$ corresponding to $E_1=0$ and $E_2\neq 0$. 

There, it was shown that the amplitude  decreases gradually according as $E_2$ increases from $0$ to a critical value $E_c$, at which $K_g=0$; and, the amplitude vanishes only at $E_2=E_c$. Since the $K_g$ is an  effective coupling constant for the $\|$ degrees of freedoms in the bi-local systems, the $\|$ coordinates of respective particles of the bi-local systems are classically free for $K_g=0$. If the bi-local systems is embedded in two-dimensional $(\bm{e}^0,\bm{e}^1)$ spacetime, the $K_g=0$ will imply the dissociation of the bi-local systems. As shown in the quantized theories, however, there arises no transition between the ground state of a free bi-local system $(E_1=0,K_g=\kappa)$ and that of a critical bi-local system with $(E_2=E_c,K_g=0)$. Thus,  the result implies that there arise no dissociation of the bi-local systems by the electric field even in such a two dimensional spacetime. 

Finally in Sec.7, the discussion has been made on the Schwinger effect of the bi-local fields corresponding to charged two particle systems. Then for $|g_{(1)}|\neq |g_{(2)}|$, we obtained a Schwinger effect similar to that of local scalar fields , with some modifications due to $\kappa\neq 0$ effect. In this case, the bi-local system describes a charged particle in totality, and the pair creations of such bound systems are available. On the contrary, for the case of neutral bi-local systems with $|g_{(1)}|=|g_{(2)}|$, the Schwinger effect does not arise, not surprisingly (FIG.\ref{pair-creation}). The result means that the parameter $\Delta_g \,(\propto \sqrt{|g_{(1)}|}-\sqrt{|g_{(2)}|})$ plays a role analogous to the order parameter in this effect under $E\neq 0$.  
\begin{figure}
\begin{minipage}{4cm}
\includegraphics[width=4cm]{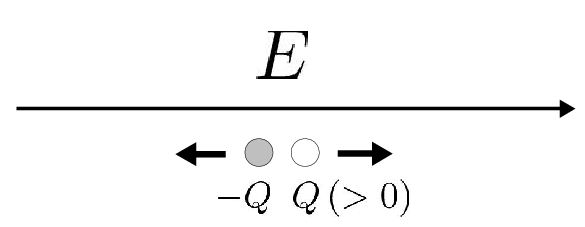}\vspace{4mm}
\end{minipage}~
\begin{minipage}{4cm}
\includegraphics[width=3.8cm]{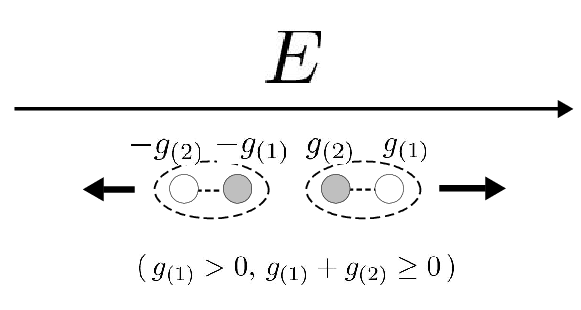}
\end{minipage}
\caption{The left side illustrates the pair creation of point like particles with charges $\pm Q$ by an electric field $E$. The right side is the case of bi-local systems consisting of charge $g_{(i)},(i=1,2)$ particles; the pair creation arises unless the total charge $g_{(1)}+g_{(2)}=0$. }
\label{pair-creation}
\end{figure}

In this paper, we have not fixed the order of $\kappa$. If we apply $\kappa$, for example, to the orders $c\sqrt{\hbar\kappa}\gtrsim $ GeV, TeV,$\cdots$, the present model will describe the mesons, the bound states of elementary particles, and so on. Conversely, if we consider the $c\sqrt{\hbar\kappa}$ as a lower atomic energy scale, then the present model will be applicable  to the problem of tunneling ionization of atoms by electric field\cite{Tunneling ionization,Dissociative tunneling}. In such a case, we have to deal with a non-uniform electric field in some way instead of a constant electric field. In this paper, we have discussed the bi-local systems with spinless particles; the extension of the bi-local systems to those with spinning particles is not difficult\cite{Casalbuoni,Crater}. Those extended applications are interesting next problems. 

\section*{Acknowledgments}

The authors wish to thank the the College of Science and Technology, Nihon University where the paper was written, for   the research support. 

\appendix

\section{The $\hat{\Pi}_{[g]\|}$ in terms of the canonical variables $(X,\hat{P};x,\hat{p})$}

In Sec.3, the canonical variables  $\{\hat{\Pi}_{[g]\|}\}$ have been introduced by an algebraic way. It will be useful to represent those variables by means of the usual center of mass and the relative canonical variables for clear understanding.

Now, with $(\epsilon_1,\epsilon_2)=(1,-1)$, we can write $\hat{\pi}_{(i)}$ as
\begin{align}
 \hat{\pi}_{(i)}=\left\{\left(\frac{1}{2}\hat{P}+\epsilon_i\hat{p}\right)+\frac{g_{(i)}E}{2c}\left(\frac{1}{2}X+\epsilon_ix\right)\right\}
 \end{align}
remembering $(\hat{p}_{(i)},x_{(i)})=\left(\frac{i}{2}\hat{P}+\epsilon_i\hat{p},X+\epsilon_i\frac{1}{2}x\right)$. Then Eq.(\ref{pi(i)}) yields
\begin{align}
 \hat{\Pi}_{g\|} &=\left(\sqrt{|\tilde{g}_{(2)}|}\hat{\pi}_{(1)}+\sqrt{|\tilde{g}_{(1)}|}\hat{\pi}_{(2)}\right)_{\|} \nonumber \\
 &=w_g\hat{P}_{\|}-\Delta_g\hat{p}_{\|}+\frac{\tilde{g}_*gE}{2c}\left(\Delta_g\tilde{X}+w_g\tilde{x}\right)_{\|}, \\
 \hat{\pi}_{g\|} &=\frac{1}{2}\left(\sqrt{|\tilde{g}_{(2)}|}\hat{\pi}_{(1)}-\sqrt{|\tilde{g}_{(1)}|}\hat{\pi}_{(2)}\right)_{\|}  \nonumber \\
 &=-\frac{\Delta_g}{4}\hat{P}_{\|}+w_g\hat{p}_{\|}+\frac{\tilde{g}_*gE}{2c}\left(w_g\tilde{X}+\frac{\Delta_g}{4}\tilde{x}\right)_{\|} .
\end{align}
With the aid of these equations, one can write the right-hand side of Eq.(\ref{Pi[g]}) as
\begin{align}
 \hat{\Pi}_{[g]\|} &=\hat{\Pi}_{g\|}-\frac{\tilde{g}_*gE}{c}\frac{\tilde{g}_*}{w_g}\tilde{x}_{\|}+\frac{\Delta_g}{w_g}\hat{\pi}_{g\|}  \nonumber \\
 &=\frac{\tilde{g}_*}{w_g}\hat{P}_{\|}+\frac{\tilde{g}_*gE}{2c}\left(2\Delta_g\tilde{X}_{\|}+\frac{1-2\tilde{g}_*}{w_g}\tilde{x}_{\|}\right) . \label{Pi[g]PX}
\end{align}
The commutation relation (\ref{Pi[g]-Pi[g]}) is a direct result of the rightest side of this representation provided $\Delta_g\neq 0$. In this case, there exits unitary transformation $\hat{U}_\Delta=\hat{U}_2\hat{U}_1$ with
\begin{align}
 \hat{U}_1 &=e^{\frac{i}{\hbar}\frac{w_g}{\tilde{g}_*}\left(X^1\frac{\tilde{g}_*gE}{2c}\right)\left(2\Delta_gX^0+\frac{1-2\tilde{g}_*}{w_g}x^0\right)},\\
\hat{U}_2 &=e^{-\frac{i}{\hbar}\frac{w_g}{\tilde{g}_*}\hat{P}^1\left(\frac{c}{2\tilde{g}_*gE\Delta_g}\hat{P}^1\right)},
\end{align}
which leads to
\begin{align}
 \hat{U}_\Delta \mathcal{P}\hat{U}_\Delta^\dag=\frac{\tilde{g}_*}{w_g}\hat{P}^1~~~\mbox{and}~~~\hat{U}_\Delta \mathcal{X}\hat{U}_\Delta^\dag=\frac{w_g}{\tilde{g}_*}\hat{X}^1.
\end{align}
The result allows us to write the bases (\ref{A-base}) in $\mathcal{X}$ representation so that
\begin{align}
 \phi_{(n)}(\mathcal{X})=\phi_{[n]}\left(\frac{w_g}{\tilde{g}_*}X^1\right)~\left(\, |\phi_{[n]}\rangle=\hat{U}_\Delta|\phi_{(n)}\rangle \,\right).  \label{D-function}
\end{align}
If necessary, one can write the $\phi_{(n)}(\mathcal{X})$ explicitly using Weber's D function (Parabolic
cylinder function)\cite{RHO-PRD}.

On the other hand, in the case of $\Delta_g=0$, $(\tilde{g}_*,w_g)=(1,1)$, the right-hand side of Eq.(\ref{Pi[g]PX}) becomes particularly simple; and we can find
\begin{align}
\begin{split}
 \hat{\Pi}_{[g]\|} &=\hat{P}_{\|}-\frac{gE}{2c}\tilde{x}_{\|}=\hat{U}_E \hat{P}_{\|}\hat{U}_E^{-1}, \label{specific Pi[g]}  \\
  \hat{U}_E &=e^{\frac{i}{\hbar}\frac{gE}{2c}\tilde{x}^{\|}\cdot X^{\|}} . 
\end{split}
\end{align}
The result says that the $\{\hat{\Pi}_{[g]\|}^\mu\,(\mu=0,1)\}$  in this case are commutable operators, the eigenvalues of which are the same as $\{\hat{P}^\mu\}$.

\section{SOLUTIONS FOR $\hat{\Lambda}_\|$}

To solve the system of equations (\ref{sys-eqs-1})-(\ref{sys-eqs-4}), let us put
\begin{align}
  A &=\frac{gE}{c}\frac{\Delta_g}{w_g}\tilde{g}_* , \\
  B &=\left\{4\kappa^2-\frac{\tilde{g}_*}{w_g^2}\left(\frac{gE}{c}\tilde{g}_*\right)^2\right\}=4K_g^2~( K_g>0), \label{B}
\end{align}
then using the Pauli  matrix $\sigma_1=\begin{pmatrix} 0 & 1 \\ 1 & 0\end{pmatrix}$, the equations for $(a_i)$ can be written as
\begin{align}
 k\bm{u} &=-4i\hbar A\sigma_1\bm{u}-2i\hbar w_gB\bm{v}, \\
 k\bm{v} &=8i\hbar w_g^{-1}\bm{u}-4i\hbar A\sigma_1\bm{v}.
\end{align}
The first equation gives $\bm{v}=\frac{i}{2\hbar w_gB}(k+4i\hbar A\sigma_1)\bm{u}$, and substituting this $\bm{v}$ for the second equation, we obtain 
\begin{align}
 \left[8i\hbar w_g^{-1}-\frac{i}{2\hbar w_gB}(k+4i\hbar A\sigma_1)^2\right]\bm{u}=0. \label{u-eq}
\end{align}
Putting, here, $\bm{u}^{\pm}=N\begin{pmatrix}1 \\ \pm 1\end{pmatrix}$, the $\sigma_1$ is replaced by $\sigma_1\bm{u}^{(\pm)}=\pm\bm{u}^{(\pm)}$; and so, the Eq.(\ref{u-eq}) is reduced to a quadratic equation of $k$ for $\bm{u}=\bm{u}^{(\pm)}$, which can be solved easily so that
\begin{align}
 k^{(\pm)}_\epsilon &=\epsilon 4\hbar\sqrt{B}-(\pm)4i\hbar A \nonumber \\
 &=4\hbar\left[ \epsilon 2K_g-(\pm)i\frac{gE}{c}\frac{\Delta_g}{w_g}\tilde{g}_* \right] , \label{k+-}
\end{align}
where the double sign in $k^{(\pm)}_\epsilon$ corresponds to that of $\bm{u}^{(\pm)}$, and the $\epsilon(=\pm)$ is the sign coming from the square root of $(k+(\pm)4i\hbar)^2$. Then reading $\bm{v}^{(\pm)}_\epsilon=\frac{i}{2\hbar w_gB}(k^{(\pm)}_\epsilon+(\pm)4i\hbar A)\bm{u}^{(\pm)}$, the set of solutions for $(a_i)$ can be represented as in Eq.(\ref{uvn+-}).

\section{On the ground states $|0^E_g,(v)^{\pm}\rangle\!\rangle$}
Since $[\hat{a}_g^{+},\hat{a}_g^{-}]=0$, the ground state for $\pm$ oscillator variables must be characterized by two equations
\begin{align}
\hat{a}_g^{-}|0_g^{\|}\rangle &=0,  \label{g-condition} \\
\hat{a}_g^{+}|0_g^{\|}\rangle &=0.  \label{g+condition}
\end{align}
The explicit form of $\hat{a}_g^\pm$ given in Eq.(\ref{ag-pm}) enable us to write the solution of Eq.(\ref{g-condition}) in the following form:
\begin{align}
 |0_g^{\|}\rangle=e^{-\frac{B_{-}}{\breve{B}_{-}}\hat{a}^{+\dag}\hat{a}^{-\dag}-\frac{i}{\breve{B}_{-}}\hat{a}^{+\dag}\hat{v}^{-}}|\Phi^{-}_0\rangle, \label{g-solution}
 \end{align}
where $|\Phi^{-}_0\rangle$ is a state satisfying $\hat{a}^{-}|\Phi^{-}_0\rangle=0$. Then, applying Eq.(\ref{g+condition}) to the state (\ref{g-solution}), one can find
\begin{align}
 |\Phi^{-}_0\rangle=e^{-\frac{i}{\breve{B}_{-}}\sqrt{\frac{2\kappa}{\hbar}}\hat{v}^{+}\hat{a}^{-\dag}}|\Phi_0\rangle,
\end{align}
where $|\Phi_0\rangle$ is a state satisfying $\hat{a}^{\pm}|\Phi_0\rangle=0$. The state $|\Phi_0\rangle$ has a structure such as $|0_{\|}\rangle\otimes |\Phi_v\rangle$, where $\hat{a}^\pm|0_{\|}\rangle=0\,(\langle 0_{\|}|0_{\|}\rangle=1)$, and $|\Phi_v\rangle$ is a state, to which $\hat{v}^\pm$ act.  Since $\hat{v}^\pm$ are related to $\hat{P}^\pm$ by unitary transformations, we can construct the eigenstates of $\hat{v}^\pm$, the $|(v)^\pm\rangle\!\rangle$ in Eq.(\ref{(v)+-}),  on $|P^{+}\rangle\otimes|P^{-}\rangle$. Then choosing $|\Phi_v\rangle=|(v)^{+}\rangle\!\rangle$, we can replace the $\hat{v}^{+}$ in Eq.(\ref{g-solution}) by the eigenvalue $v^{+}$. The resultant ground state becomes the $\hat{G}|0_{\|}\rangle\otimes|(v)^{+}\rangle\!\rangle$ in Eq.(\ref{ground-1}).

The next task is to study the way of normalization of those ground states for a given $P^{\|}$ under a common $E$.  A key to this end can be found in the inner products of the states:
\begin{align}
  |\alpha,\beta.\gamma\rangle=e^{-\alpha\hat{a}^{-\dag}\hat{a}^{+\dag}-i\beta\hat{a}^{-\dag}-i\gamma\hat{a}^{+\dag}}|0_{\|}\rangle.
\end{align}
With the aid of the formula
\begin{align}
 \int\frac{d^2z}{\pi}e^{-D|z|^2+Az+Bz^*}=\frac{1}{D}e^{\frac{1}{D}AB} \\
 \left(~ {\rm Re}(D)>0,~[A,B]=0~\right), \nonumber
\end{align}
one can verify that
\begin{align}
 \langle &\alpha_1, \beta_1,\gamma_1|\alpha_2,\beta_2,\gamma_2\rangle  \nonumber \\
     &=\int\frac{d^2z}{\pi}e^{-|z|^2}\langle 0_{\|}|e^{(iz\sqrt{\alpha_1^*}+i\beta_1^*)\hat{a}^{-}+(iz^*\sqrt{\alpha_1^*}+i{\gamma}_1^*)\hat{a}^{+}} \nonumber \\
   &\times e^{-\alpha_2\hat{a}^{-\dag}\hat{a}^{+\dag}-i\beta_2\hat{a}^{-\dag}-i\gamma_2\hat{a}^{+\dag}}|0_{\|}\rangle \otimes |v^{+}\rangle  \nonumber \\
  &=\int\frac{d^2z}{\pi}e^{-\left(1-\alpha_1^*\alpha_2\right)|z|^2}e^{z\left\{\alpha_2\sqrt{\alpha_1^*}(\gamma_1^\dag)+\gamma_2\sqrt{\alpha_1^*}\right\}} \nonumber \\
  &\times e^{z^*\left\{\alpha_2\sqrt{\alpha_1^*}(\beta_1^*)+\beta_2\sqrt{\alpha_1^*} \right\}}e^{\alpha_2\beta_1^*\gamma_1^*+\beta_2\gamma_1^*+\gamma_2\beta_1^*} \nonumber \\
    &=\frac{1}{1-\alpha_1^*\alpha_2}e^{\frac{1}{1-\alpha_1^*\alpha_2}(1-\alpha_1^*\alpha_2)(\alpha_2\beta_1^*\gamma_1^*+\beta_2\gamma_1^*+\gamma_2\beta_1^*)} \nonumber  \nonumber \\
    &\times e^{\frac{1}{1-\alpha_1^*\alpha_2}\left\{\alpha_2^2\alpha_1^*(\gamma_1^*\beta_1^*)+\alpha_2\alpha_1^*(\gamma_1^*\beta_2)+\alpha_2\alpha_1^*(\gamma_2\beta_1^*)+\alpha_1^*(\gamma_2\beta_2) \right\} } \nonumber \\
     &=\frac{1}{1-\alpha_1^*\alpha_2}e^{\frac{1}{1-\alpha_1^*\alpha_2}\left\{(\alpha_2\beta_1^*+\beta_2)\gamma_1^*+(\alpha_1^*\beta_2+\beta_1^*)\gamma_2 \right\} }. \label{inner product}
  \end{align} 
Substituting $\alpha_i=\alpha=\frac{B_{-}}{\breve{B}_{-}}$, $\beta_i=\beta=\frac{1}{\breve{B}_{-}}\sqrt{\frac{2\kappa}{\hbar}}v^{-}$ and $\gamma_i=\frac{1}{\breve{B}_{+}}\sqrt{\frac{2\kappa}{\hbar}}v_i^{+}~\,(i=1,2)$ for Eq.(\ref{inner product}), we obtain
\begin{align}
 \langle \alpha,\beta,\gamma_1 |\alpha,\beta,\gamma_2\rangle  \nonumber &=\frac{1}{1-|\alpha|^2}e^{\frac{1}{1-|\alpha|^2}\left\{(\alpha\beta^*+\beta)\gamma_1^* +(\alpha^*\beta+\beta^*)\gamma_2 \right\} } \nonumber \\
  &=\frac{(1+a)^2+b^2}{4a}e^{\frac{\kappa}{\hbar a}\frac{(1+a)}{(1+a)^2+b^2}v^{-}(v_1^{+}+v_2^{+})} \nonumber \\
  &\times e^{i\frac{\kappa}{\hbar a}\frac{b}{(1+a)^2+b^2}v^{-}(v_1^{+}-v_2^{+})}. \label{intermediate}
\end{align}
The $\langle \alpha,\beta,\gamma_1 |\alpha,\beta,\gamma_2\rangle$ plays the essential role in the  inner product of $\hat{G}|0_{\|}\rangle\otimes|(v_i)^{+}\rangle\!\rangle\,(i=1,2)$; and, the real exponent in Eq.(\ref{intermediate}) brings harm for the normalization of those states with a finite norm.

Hence,  we modify the  ground states so that this real exponent factor is removed as in Eq.(\ref{modified ground state}). We also note that the definition $|(v)^\pm\rangle\!\rangle=\hat{U}^{\pm 1}|P^{\|}\rangle\!\rangle$ $\left(v^\pm=-\frac{\Delta_g}{4w_gK_g}P^\pm\right)$ with  $\hat{U}(X^{\|})=e^{\frac{i}{\hbar}\frac{4}{\Delta_g}\frac{\tilde{g}_*gEw_g}{2c}X^{-}X^{+}}$in $X$ representation leads to the completeness
\begin{align}
 \int d^2P^{\|}|(v)^\pm\rangle\!\rangle\langle\!\langle (v)^\pm|=\int d^2P^{\|}|P^{\|}\rangle\!\rangle\langle\!\langle P^{\|}|=1. \label{complete}
 \end{align}
Then with the help of Eq.(\ref{complete}), we obtain
\begin{align}
 I_{12} &\equiv \langle\!\langle 0^E_g,(v_1)^{+}|0^E_g,(v_2)^{+}\rangle\!\rangle  \nonumber \\
 &= \int d^2P^{\|}\langle\!\langle 0^E_g,(v_1)^{+}|(v)^{-}\rangle\!\rangle\langle\!\langle (v)^{-}|0^E_g,(v_2)^{+}\rangle\!\rangle \nonumber \\
 &=\int d^2P^{\|}N_{v^{-},v_1^{+}}N_{v^{-},v_2^{+}}\langle \alpha,\beta,\gamma_1 |\alpha,\beta,\gamma_2\rangle \nonumber \\
 &\times \langle\!\langle (v_1)^{+}|(v)^{-}\rangle\!\rangle\langle\!\langle (v)^{-}|(v_2)^{+}\rangle\!\rangle \nonumber \\
 &=\left|\frac{(1+a)^2-b^2}{(1+a)^2+b^2}\right|\int d^2P^{\|}e^{i\frac{\kappa}{\hbar a}\frac{b}{(1+a)^2+b^2}v^{-}v_{12}^{+} } \nonumber \\
 &\times \langle\!\langle (v_1)^{+}|(v)^{-}\rangle\!\rangle\langle\!\langle (v)^{-}|(v_2)^{+}\rangle\!\rangle. \label{inner-1}
\end{align}
where $\beta=\beta(v^{-}),\gamma_i=\gamma(v_i^{+})$ and $v_{ij}^\pm=v_i^\pm-v_j^\pm$.
Here, the inner products $\langle\!\langle(v_i)^{\mp}|(v_j)^{\pm}\rangle\!\rangle=\langle\!\langle P_i^{\|}|\hat{U}^{\pm 2}|P_j^{\|}\rangle\!\rangle$ are specific cases ($n=\pm 2$) of 
\begin{align}
 \langle\!\langle P_i^{\|}|&\hat{U}^n| P_j^{\|}\rangle\!\rangle=\int d^2X^{\|}\langle\!\langle P_i^{\|}|X^{\|}\rangle\!\rangle\hat{U}^n(X^{\|})\langle\!\langle X^{\|}|P_j^{\|}\rangle\!\rangle \nonumber \\
 &=\frac{1}{n}\left(\frac{\Delta_g}{4w_g}\frac{2c}{\tilde{g}_*gE}\right)\frac{1}{2\pi\hbar}e^{-\frac{i}{\hbar}\frac{1}{n}\left(\frac{\Delta_g}{4w_g}\frac{2c}{\tilde{g}_*gE}\right)P_{ij}^{-}P_{ij}^{+} }. \label{inner-2}
\end{align}
Applying this formula to Eq.(\ref{inner-1}), and remembering $\frac{\kappa}{2ab}v^{-}v^{+}=\left(\frac{\Delta_g}{4w_g}\frac{2c}{\tilde{g}_*gE}\right)P^{-}P^{+}$, we have
\begin{align}
 I_{12} &=\left|\frac{(1+a)^2-b^2}{(1+a)^2+b^2}\right|\int d^2P_i^{\|} \left\{\frac{1}{2}\left(\frac{\Delta_g}{4w_g}\frac{2c}{\tilde{g}_*gE}\right)\frac{1}{2\pi\hbar}\right\}^2 \nonumber \\
 &\times e^{i\frac{\kappa}{\hbar a}\frac{2b^2}{(1+a)^2+b^2}\left(\frac{\Delta_g}{4w_g}\frac{2c}{\tilde{g}_*gE}\right)P_i^{-}P_{12}^{+} } \nonumber \\
 &\times e^{\frac{i}{\hbar}\frac{1}{2}\left(\frac{\Delta_g}{4w_g}\frac{2c}{\tilde{g}_*gE}\right)\left(P_{i1}^{-}P_{i1}^{+}-P_{i2}^{-}P_{i2}^{+}\right) }. \label{inner-2}
 \end{align}
 Thus, because of  ${\small (P_{i1}^{-}P_{i1}^{+}-P_{i2}^{-}P_{i2}^{+})=(P_1^{-}P_1^{+}-P_2^{-}P_2^{+}) }$ ${\small -P_i^{-}P_{12}^{+}-P_i^{+}P_{12}^{-}}$, we finally arrive at
\begin{align}
 \langle\!\langle 0^E_g &,(v_1)^{+}|0^E_g,(v_2)^{+}\rangle\!\rangle=\delta^{(2)}\left(P_1^{\|}-P_2^{\|}\right).
\end{align}

Next, we examine the inner products between the state $|\emptyset,{\rm v}_1\rangle\!\rangle$, the ground state for $E=0$, and the state $|0^E_g,(v_2)^{+}\rangle\!\rangle$, the ground state for $E\neq 0$. In this case, since $|\emptyset,(v)^{+}\rangle\!\rangle$ does not contain $\hat{a}^{-\dag}\hat{a}^{+\dag}$, we can evaluate the inner product in such a way that
\begin{align}
 I_{12}^{\emptyset} &\equiv  \langle\!\langle \emptyset,{\rm v}_1|0_g,(v_2)^{+}\rangle\!\rangle \nonumber \\
 &=e^{-\frac{\kappa}{2\hbar}{\rm v}_1^{-}{\rm v}_1^{+}}\langle 0^{\|}|e^{\frac{i}{2}\sqrt{\frac{2\kappa}{\hbar}}{\rm v}_1^{+}\hat{a}^{-}}e^{\frac{i}{2}\sqrt{\frac{2\kappa}{\hbar}}{\rm v}_1^{-}\hat{a}^{+}} \nonumber \\
  &\times \int d^2P_i^{\|}e^{-\alpha\hat{a}^{+\dag}\hat{a}^{-\dag}-i{\beta}(v_i^{-})\hat{a}^{+\dag}}e^{-i{\gamma}({v}_2^{+})\hat{a}^{-\dag}}|0^{\|}\rangle \nonumber \\
  &\times N_{{v}^{-},v_2^{+}}\langle\!\langle P_1^{\|}|(v_i)^{-}\rangle\!\rangle\langle\!\langle (v_i)^{-}|(v_2)^{+}\rangle\!\rangle  \nonumber \\
  &=e^{-\frac{\kappa}{2\hbar}{\rm v}_1^{-}{\rm v}_1^{+}+\alpha\left(\frac{\kappa}{2\hbar}\right){\rm v}_1^{-}{\rm v}_1^{+}}e^{\frac{1}{\breve{B}_{+}}\frac{\kappa}{\hbar}{v}_2^{+}{\rm v}_1^{-}}\!\int d^2P_i^{\|}e^{\frac{1}{\breve{B}_{-}}\frac{\kappa}{\hbar}v_i^{-}{\rm v}_1^{+}} \nonumber \\
  &\times N_{v^{-},v_2^{+}}\langle\!\langle P_1^{\|}|(v_i)^{-}\rangle\!\rangle\langle\!\langle (v_i)^{-}|(v)_2^{+}\rangle\!\rangle .
\end{align}
Here, $\langle\!\langle P_1^{\|}|(v)^{-}\rangle\!\rangle$ and $\langle\!\langle (v)^{-}|(v)_2^{+}\rangle\!\rangle$ are specific cases of Eq.(\ref{inner-2}) corresponding to $n=-1$ and $n=2$ respectively, and $N_{v^{-},v_2^{+}}=N_0e^{-\frac{\kappa}{\hbar a}\frac{(1+a)}{(1+a)^2+b^2}v^{-}v_2^{+}}$. Thus,
 \begin{align}
  I_{12}^{\emptyset} &=N_0e^{-\frac{\kappa}{2\hbar}{\rm v}_1^{-}{\rm v}_1^{+}+\alpha\left(\frac{\kappa}{2\hbar}\right){\rm v}_1^{-}{\rm v}_1^{+}}e^{\frac{1}{\breve{B}_{+}}\frac{\kappa}{\hbar}{v}_2^{+}{\rm v}_1^{-}} \nonumber \\
 &\times \int d^2P_i^{\|}e^{\frac{1}{\breve{B}_{-}}\frac{\kappa}{\hbar}{v}_i^{-}{\rm v}_1^{+}}  e^{-\frac{\kappa}{\hbar}\frac{(1+a)}{a\{(1+a)^2+b^2\}}v_i^{-}v_2^{+}} \nonumber \\
 &\times 2\left\{\left(\frac{\Delta_g}{4w_gK_g}\right)^2\frac{1}{2\pi\hbar}\frac{\kappa}{2ab}\right\}^2e^{\frac{i}{\hbar}\frac{\kappa}{2ab}\left(2v_{1i}^{-}v_{1i}^{+}-v_{i2}^{-}v_{i2}^{+}\right)}.
\end{align}
Remembering further $\left(\frac{\Delta_g}{4w_gK_g}\right)^2d^2P^{\|}=d^2v^\pm$, the integration can be carried out; and, we arrive at the representation of $I_{12}^{\emptyset}$ in Eq.(\ref{transition}):
 \begin{align}
  I_{12}^{\emptyset} &=N_0e^{-\frac{\kappa}{2\hbar}{\rm v}_1^{-}{\rm v}_1^{+}+\alpha\left(\frac{\kappa}{2\hbar}\right){\rm v}_1^{-}{\rm v}_1^{+}}e^{\frac{1}{\breve{B}_{+}}\frac{\kappa}{\hbar}{v}_2^{+}{\rm v}_1^{-}} \nonumber \\
 &\times  e^{\frac{\kappa}{\hbar}(2v_1^{-}-v_2^{-})\left(\frac{1}{\breve{B}_{-}}v_1^{+}-\frac{(1+a)}{a\{(1+a)^2+b^2\}}v_2^{+}\right) } \nonumber \\
  &\times  \left(\frac{\Delta_g}{4w_gK_g}\right)^2\left(\frac{1}{2\pi\hbar}\frac{\kappa}{ab}\right)e^{-\frac{i}{\hbar}\frac{\kappa}{ab}v_{12}^{-}v_{12}^{+}}. \label{I_12_emptyset}
\end{align}
It should be noticed that by virtue of
\begin{align}
 \lim_{b\rightarrow 0} \left(\frac{1}{2\pi\hbar}\frac{\kappa}{ab}\right)e^{-\frac{i}{\hbar}\frac{\kappa}{ab}v_{12}^{-}v_{12}^{+}}=\delta(v_{12}^{-})\delta(v_{12}^{+}),
 \end{align}
one can find the limit $I_{12}^{\emptyset}\rightarrow \delta^{(2)}(P^{\|})\, (E\rightarrow 0)$.

\end{document}